\documentclass{article}

\usepackage{PRIMEarxiv}

\usepackage[utf8]{inputenc} 
\usepackage[T1]{fontenc}    
\usepackage[hidelinks]{hyperref}
\usepackage{url}            
\usepackage{booktabs}       
\usepackage{amsfonts}       
\usepackage{nicefrac}       
\usepackage{microtype}      
\usepackage{lipsum}
\usepackage{fancyhdr}       
\usepackage{graphicx}       
\graphicspath{{media/}}     

\usepackage{amsmath,amssymb,amsfonts}
\usepackage{algorithmic}
\usepackage{textcomp}

\usepackage{subfiles}
\usepackage{subcaption}
\usepackage{rotating}
\usepackage{adjustbox}
\usepackage[table,xcdraw]{xcolor}
\usepackage{lscape}

\title{Towards Fine-Grained Localization of Privacy Behaviors}

\author{
  Vijayanta Jain, \\
  University of Maine \\
  Orono, ME\\
  \texttt{vijayanta.jain@maine.edu} \\
  \And
  Sepideh Ghanavati\\
  University of Maine \\
  Orono, ME\\
  \texttt{sepideh.ghanavati@maine.edu} \\
  \And
  Sai Teja Peddinti \\
  Google Inc. \\
  Mountain View, CA\\
  \texttt{psaiteja@google.com} \\
  \And
  Collin McMillan \\
  University of Notre Dame \\
  Notre Dame, IN \\
  cmc@nd.edu \\ }

\begin{document}
\maketitle

\begin{abstract}
Mobile applications are required to give privacy notices to the users when they collect or share personal information. Creating consistent and concise privacy notices can be a challenging task for developers. Previous work has attempted to help developers create privacy notices through a questionnaire or predefined templates. In this paper, we propose a novel approach and a framework, called PriGen, that extends these prior work. PriGen uses static analysis to identify Android applications' code segments which process sensitive information (i.e. \textit{permission-requiring code segments}) and then leverages a Neural Machine Translation model to translate them into \textit{privacy captions}. We present the initial evaluation of our translation task for $\sim$300,000 code segments.
\end{abstract}


\section{Introduction}

Privacy notices describe \textit{how} an application uses personal information and \textit{why} (i.e., its privacy behaviors), and help users make informed privacy decisions. Application developers are required by privacy regulations \cite{GDPR, CCPA} and application store policies \cite{google-play} to provide users with accurate privacy notices. Recently, both application stores (i.e., App Store \cite{AppStoreLabels} and Google Play \cite{PlayLabels}) introduced their versions of privacy ``nutrition'' labels (or simply privacy labels) \cite{kelley2009nutrition} to simplify this process. These labels are standardized notice formats that help developers easily describe \textit{how} and \textit{why} their application uses personal information and build trust with the users \cite{li2022Understanding}. 

Existing challenges in creating privacy \textit{notices} hinder providing accurate \textit{labels}, as well. These challenges range from, difficulty in comprehending privacy behaviors of their applications \cite{peddinti2019reducing, li2022Understanding}, to gaps in developers' knowledge about privacy concepts \cite{hadar2018privacy}. Failure to provide accurate labels violates privacy regulations, which can result in hefty fines for developers \cite{GDPR-Fines}. These inaccuracies can also impact users' well-being since it inhibits their ability in making privacy-preserving decisions. Lastly, discrepancies between labels and applications' privacy behaviors also diminish trust between developers and users. 

Recent works aim to address the challenges of generating privacy notices, including privacy labels \cite{yu2016toward, zimmeck2021privacyflash, li2021honeysuckle, gardnerhelping2022, jain2022pact}. Some of these efforts leverage static analysis approaches to identify APIs called and use templates \cite{yu2016toward}, questionnaires \cite{zimmeck2021privacyflash}, or developers' annotations \cite{li2021honeysuckle} to generate notices; while others use machine learning (ML) approaches~\cite{jain2021prigen, jain2022pact}. For example, Gardner et al. \cite{gardnerhelping2022} develop \textit{Privacy Label Wiz} which uses static analysis to analyze iOS applications' source code, provides the summary of the results to developers, and prompts them with questions to help create privacy labels for the App Store. Jain et al. \cite{jain2022pact} create platform-neutral \textit{Privacy Action} labels that describe \textit{how} and \textit{why} an application's code uses personal information. They use deep learning to predict these labels from the source code. 

\begin{figure*}[h!]
    \centering
    \fbox{\includegraphics[width=0.95\textwidth]{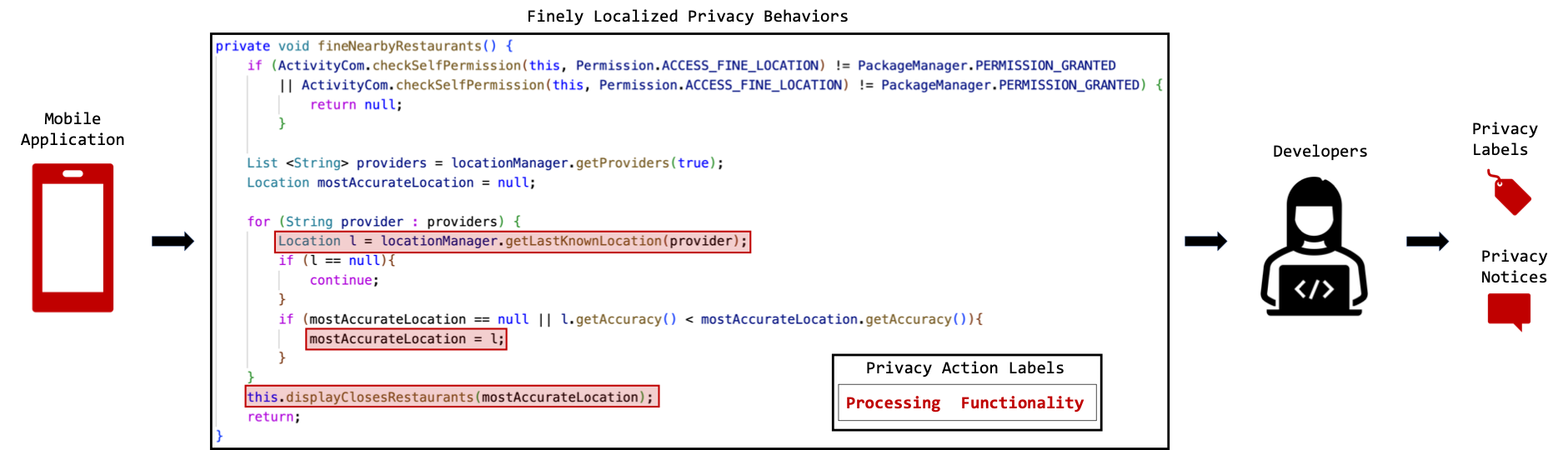}}
    \caption{Example of how fine-grained localization can help developers create accurate privacy labels and notices. 
    }
    \label{fig:source-code-example}
\end{figure*}

While these approaches aid developers in creating privacy labels or notices, they have several drawbacks. For example, Privacy Label Wiz \cite{gardnerhelping2022} helps understand the privacy behaviors of applications with analysis summary and prompts, but it does not automate the process of creating labels or identifying the purposes. Label creation and purpose identification are still the developers' responsibilities. Moreover, as the application evolves, the tool will rely on developers to keep track of changes in privacy behaviors, including purpose, which may increase the developers' effort. This practice can be especially challenging in settings where developers are part of large teams \cite{li2022Understanding} and the rationale for using personal information is distributed among members. Jain et al. \cite{jain2022pact} address some of these limitations by automating the process of creating \textit{Privacy Action} labels. However, their approach lacks the necessary source code context which makes it challenging for developers to understand the privacy behaviors of their source code and create privacy notices \textit{solely based on labels}. Consider the code snippet in Figure~\ref{fig:source-code-example}. If developers are only provided with the \textit{Privacy Action} labels \texttt{Processing} and \texttt{Functionality}, they must read the method to comprehend its behavior and understand how the personal information is \texttt{Processed} and for which \texttt{Functionality} it is used. The complexity of this task increases when the length of code snippets increases and it spans to include multiple methods/classes.

In this paper, we propose \textit{fine-grained localization} of privacy behaviors.  In fine-grained localization, we locate individual statements in source code that encode privacy behaviors and use them to predict their privacy labels. These localized statements together with predicted privacy labels can then help developers create short sentences to be used in privacy notices. The key novelty of our approach lies in the granularity of locating where privacy behaviors are implemented in source code. While previous localization approaches limit localization to only classes or methods \cite{ma2020droidetec, narayanan2018multi, wu2021android} (i.e., coarse-grained localization), our approach moves one step further by also identifying which \textit{statements} within a method implement privacy behaviors (i.e., fine-grained localization). Similarly, static analysis approaches \cite{zimmeck2021privacyflash, gardnerhelping2022} only rely on specific types of statements in a single method, such as API calls, to create privacy notices. Our approach identifies not only API calls but also other statements across multiple methods that use personal information to create privacy labels. Apart from the novelty, our approach provides the benefits of previous efforts such as helping developers understand high-level privacy behaviors (similar to \textit{Privacy Label Wiz} \cite{gardnerhelping2022}) as well as fine-grained ones and generating privacy labels (similar to Jain et al.\cite{jain2022pact}) but with higher accuracy (at least $\sim$11\% higher). Our approach is also complementary to existing static and dynamic analysis approaches and provides an \textit{alternative technique} to generate privacy labels. By identifying multiple methods and employing machine learning, our fine-grained localization approach identifies individual statements in those methods providing a much more granular view of privacy behaviors implemented in the application's source code.


We explain how our approach can help create accurate labels and privacy notices with the following example. Consider the same code snippet in Figure \ref{fig:source-code-example}: If developers are provided with the localized statements (i.e., segments highlighted with red boxes), \textit{along} with the predicted labels of \texttt{Processing} and \texttt{Functionality}, they can better understand \textit{how location is processed} (i.e., calculating distance) and \textit{for what functionality} (i.e., suggest nearby restaurants) in source code. To provide labels to users, developers can use these predicted labels (i.e., \texttt{Processing} and \texttt{Functionality}) and map them to their platform-specific labels (App Store or Google Play). They can also use the labels and localized statements to create privacy statements such as ``We use your location to calculate the distance and suggest nearby restaurants.'' and use them in their permission rationales or privacy policies. 



To implement our approach, we develop a multi-head encoder model that creates individual representations of multiple methods and uses `attention' \cite{vaswani2017attention} to identify relevant statements in those methods.\footnote{A recent NLP work has raised questions about an attention module's capability to identify relevant parts of input sequences \cite{jain2019attention}. However, several studies indicate that attention is important \cite{serrano2019attention, sun2020understanding} and can be used for interpreting the results of a classification task. Hence, we use attention to select relevant statements and localize privacy behaviors.}  We demonstrate the efficacy of our approach by training the model on the publicly released ADPAc\footnote{\url{https://github.com/PERC-Lab/PAcT}} dataset~\cite{jain2022pact}, which contains source code samples and their \textit{Privacy Action} labels. We choose the components of our model by conducting six sets of experiments. In each set, we train a classifier with 24 datasets in the ADPAc, evaluate the optimal model and dataset configurations, and use them for our model. We also analyze the results and provide key insights into how to choose the best combination of model and dataset configurations for identifying privacy behaviors in code.

We evaluate our work both qualitatively and quantitatively. For qualitative evaluation, we recruit six software professionals, with experience in software development and privacy-related research. We ask them to write simple sentences that describe privacy behaviors for code samples with and without fine-grained localization. Our evaluation shows that while there are negligible differences in the statements that the professionals write with/out localization (since they all have privacy expertise), the time and mental effort required are significantly less when the code samples are finely localized. Furthermore, professionals with less experience benefit the most from localization, since they save up to $\sim$74\% of time to write statements of comparable quality and details to those written by the most experienced ones. Quantitatively, we use accuracy and F-1 scores to gauge the model's performance in predicting \textit{Privacy Action} labels. Our evaluation shows our model increases the baseline accuracy \cite{jain2022pact} by at least 11\% with up to 30\% for some labels. Our lowest accuracy is 91.41\% and the highest is 98.45\% across labels. We also measure the accuracy of fine-grained localization by asking three of the six software professionals who are more experienced in software development and privacy to manually inspect the statements in code samples that are highlighted by the attention module. The results show that at least one annotator agrees that 85\% of statements implement privacy behaviors. Our analysis strongly demonstrates that our approach can identify privacy labels and help in writing high-quality privacy statements. 


In summary, our main contributions are as follows:

\begin{enumerate}
    \item We address the issue of creating accurate privacy labels by developing automated fine-grained localization to identify statements in methods that implement privacy behaviors and that explain \textit{how} and \textit{why} personal information is used. This approach extends the granularity of localization in previous works to individual statements.
    \item We implement our approach by developing a novel attention-based deep learning model, for which we conducted six sets of experiments with 24 datasets to meticulously choose the best model and dataset configurations. We share insights from these experiments to help researchers use them as a blueprint for classifying privacy behaviors in source code. 
    \item Our evaluation demonstrates that our approach establishes a new state-of-the-art for predicting \textit{Privacy Action} labels, its efficacy in finely localizing privacy behaviors, and helping create privacy statements by reducing time and effort. 
\end{enumerate}

\label{sec:introduction}

\section{Related Work}

\begin{figure}[!t]
	\centering
\includegraphics[width=7cm]{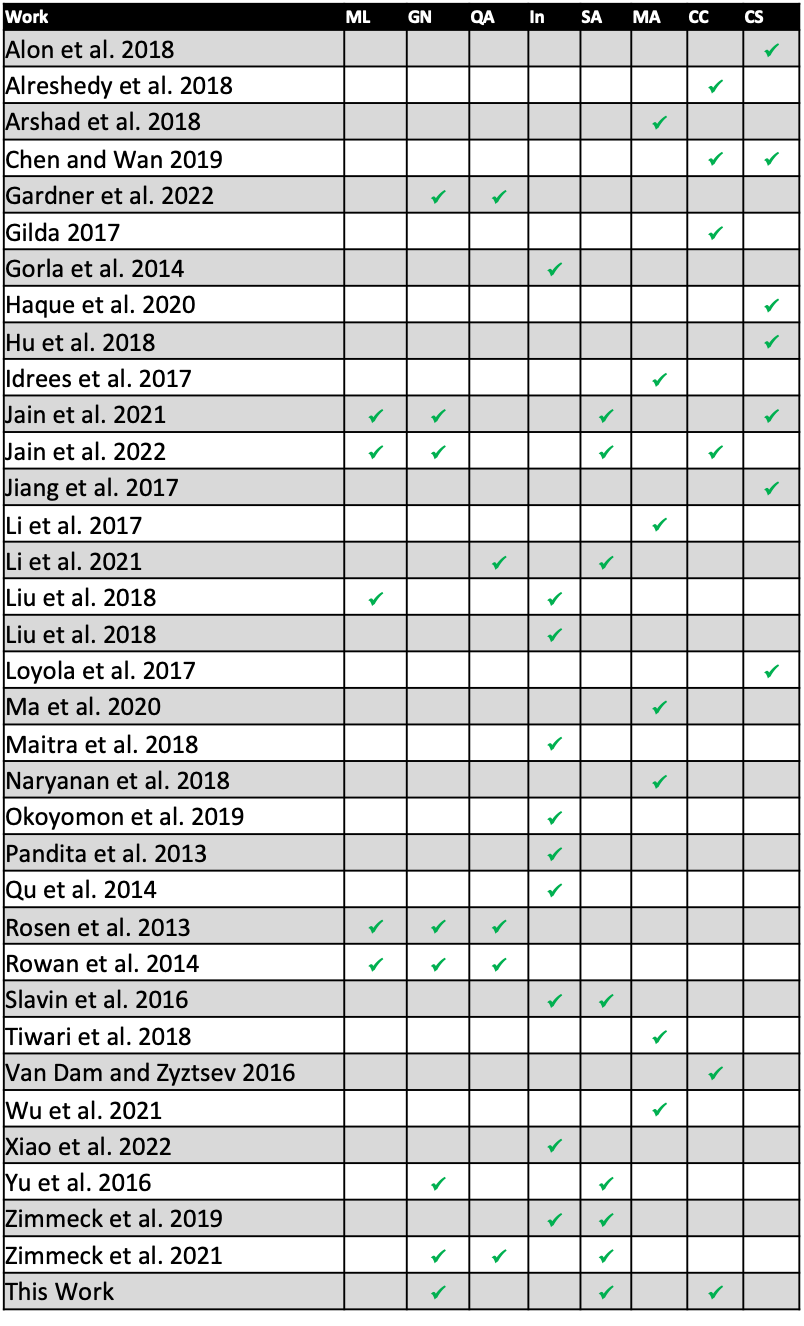}
	\caption{Selection of closely-related, peer-reviewed publications. Column \textit{`ML'} =  ML based; \textit{`GN'} = Generate Notices; \textit{`Q/A'} Question-Answering based; \textit{`In'} = Inconsistency Analysis; \textit{`SA'} = Static/Dynamic Analysis; \textit{`MA'} = Malicious Applications; \textit{`CC'} = Code Classification and \textit{`CS'} = Code Summarization.}
	\label{fig:related-work}
\end{figure}

Figure \ref{fig:related-work} summarizes the prior work by showcasing trends in finding discrepancies between privacy notices and application source code, creating privacy notices, identifying malicious applications, and classifying and summarizing code segments. 

\subsection{Finding Discrepancies}

A significant effort has been made to identify discrepancies between an application's privacy behaviors and its privacy notices~\cite{zimmeck2019maps, gorla2014checking, liu2018large, slavin2016pvdetector, okoyomon2019ridiculousness, maitra2018privacy, xiao2022lalaine}. Most works focus on matching the application's privacy behaviors with its privacy policy \cite{zimmeck2019maps, maitra2018privacy}, application description~\cite{gorla2014checking, qu2014autocog, pandita2013whyper}, permission rationales \cite{liu2018large}, or privacy labels \cite{xiao2022lalaine}, using static analysis, and counting the instances of discrepancies. These works have identified significant inconsistencies between applications' privacy behaviors and their notices and \textit{highlighted} the issue. Our work aims at \textit{resolving} these discrepancies by generating accurate privacy labels and localizing privacy behaviors in source code to help create consistent and detailed notices. 

\subsection{Creating Privacy Notices}

A number of studies focus on creating privacy notices \cite{rowan2014encouraging, gardnerhelping2022, rosen2013appprofiler, yu2016toward, zimmeck2021privacyflash, li2021honeysuckle, jain2021prigen, jain2022pact}. These approaches either create privacy policies using questionnaires \cite{rowan2014encouraging, rosen2013appprofiler, zimmeck2021privacyflash} (similar to privacy policy generators), or provide notices in different formats, such as permission rationales \cite{liu2018mining}, privacy describing statements \cite{yu2016toward},  or in-application notices \cite{li2021honeysuckle}. AutoPPG \cite{yu2016toward}, PrivacyFlash Pro \cite{zimmeck2021privacyflash}, Privacy Label Wiz \cite{gardnerhelping2022}, Honeysuckle \cite{li2021honeysuckle}, and PAcT \cite{jain2022pact} are closely related to our work since they also aim to aid developers to create privacy notices using source code. AutoPPG \cite{yu2016toward} analyzes the APIs called to identify the personal information used and then uses a static \texttt{subject form object [condition]} template to create privacy statements, but it lacks the rationale for using personal information. PrivacyFlash Pro \cite{zimmeck2021privacyflash}, Honeysuckle \cite{li2021honeysuckle}, and Privacy Label Wiz \cite{gardnerhelping2022} provide rationales in their notices, however, they rely on developers' efforts to do so. Jain et al. \cite{jain2022pact} automatically provide \textit{Privacy Action} labels that describe \textit{how} and \textit{why} personal information is used. However, as discussed in Section \ref{sec:introduction}, these labels lack source code context; adding more work for developers to understand their privacy behaviors. In this paper, we extend these efforts, by automating the process of creating accurate privacy labels and providing developers with source code context. These labels and context will help developers understand the privacy behaviors of their applications.

\subsection{Identifying Malicious Applications}

Studies that identify malicious mobile applications are tangential to our work; however, there are some similarities in our approaches that are worth mentioning. Several works have provided approaches to identify malicious applications \cite{arshad2018samadroid, idrees2017pindroid, tiwari2018android} which pose security risks. Recently, some studies extended these efforts to localize malicious code in applications \cite{ma2020droidetec, narayanan2018multi, wu2021android, li2017locating}. These approaches use program graphs, such as call or dependency graphs, to represent an application, extract features, and classify them using an attention-based deep learning model \cite{ma2020droidetec, narayanan2018multi, wu2021android}. Using attention weights, these approaches identify relevant features used for classification, which they use to localize malicious source code. Our approach to localizing privacy behaviors in source code follows a similar logic. However, a significant difference between our approaches is the granularity of localization. These approaches limit localization to malicious packages \cite{li2017locating} or methods \cite{ma2020droidetec, narayanan2018multi, wu2021android} (i.e., coarse-grained localization), whereas, in our approach, we localize statements within a method which provides a more fine-grained localization of privacy behaviors. 

\subsection{Classifying and Summarizing Code} 

Our work also draws inspiration from efforts in software engineering to classify and summarize code. Classification tasks in this field primarily focus on detecting the programming language of code snippets \cite{alreshedy2018SCC, gilda2017source, van2016software}, whereas summarization tasks focus on transforming code snippets into natural language text for various tasks. The format of the generated text differs based on the purpose, such as transforming code differences into commit messages for version control \cite{jiang2017automatically, loyola2017neural}, or into comments for documentation \cite{alon2018code2seq, hu2018deep, haque2020improved, chen2019neural}. Our work extends these efforts to comprehend code and modifies it to finely localize privacy behaviors and predict their privacy labels.

\label{sec:related-work}

\section{Background}


\subsection{Permissions and Static Analysis}
\label{subsec:static-analysis}

Mobile operating systems, such as Android, employ a permissions system to allow developers to access users' personal information while giving control to users to protect their personal information. In this system, if developers want to access personal information, they call a system API \footnote{\url{https://developer.android.com/reference/}} and declare necessary permissions, such as in \texttt{AndroidManifest.xml} file for Android applications. When the application requires access to sensitive information, users are asked if they would permit the application to access this information. In case they agree, users give control of their personal information. 

Previous work uses static analysis to identify system API calls and extract methods that call them \cite{jain2022pact}. They refer to these extracted code snippets as \textit{Permission-Requiring} Code Segments (PRCS) since they call APIs that require permission (i.e., \textit{permission-requiring} APIs). Since a method calling permission-requiring APIs can share the accessed information with other methods for further use, a PRCS includes multiple methods linked via a call graph. Each PRCS segment contains at most three methods because they found that in $\sim$80\% of cases, personal information is used within these three methods \cite{jain2021prigen, jain2022pact}. Since personal information can ``hop'' between methods in a PRCS, each method is referred to as a ``hop''. The first hop in a PRCS is the method that calls the permission-requiring API and each PRCS contains at least this first hop. The subsequent methods are called the second and third hops, respectively (we interchangeably use ``hop'' and ``method''). In our approach, we predict labels and localize privacy behaviors for PRCS that consist of up to three hops. Applications can also access and use personal information via user interfaces \cite{nan2015uipicker, andow2017uiref, huang2015supor}. However, in this work, we limit the scope of code segments to the ones that call system APIs, since our goal is to demonstrate the feasibility of fine-grained localization and not to show the coverage of extracting source code.

\subsection{Privacy Action Taxonomy and Dataset}
\label{subsec:pact-labels}

Privacy Action Taxonomy (PAcT) is a taxonomy that defines privacy behaviors implemented in source code \cite{jain2022pact}. The goal of this taxonomy is to help consistently detect privacy behaviors in the application's source code and create privacy labels for them. In this taxonomy, there are two categories of labels: \textit{Practice} and \textit{Purpose}. The labels in the \textit{Practice} category describe \textit{how} a code segment uses personal information whereas the labels in the \textit{Purpose} category answer \textit{why}. Both categories contain 4 labels each. The \textit{Practice} category contains \texttt{Processing}, \texttt{Collecting}, \texttt{Sharing}, and \texttt{Other} labels. Whereas the \textit{Purpose} category contains \texttt{Functionality}, \texttt{Advertisement}, \texttt{Analytics}, and \texttt{Other} labels. The definition of each label is described in Jain et al.\cite{jain2022pact}.

Jain et al. \cite{jain2022pact} used PAcT to create an annotated dataset (called ADPAc)  of code segments and their \textit{Privacy Action} labels. This dataset contains $\sim$5,200 PRCS and $\sim$14,000 labels, which is publicly available (see Section~\ref{sec:introduction}). Since each code segment can implement multiple behaviors, some samples are annotated with multiple labels; hence, ADPAc is a multi-class multi-label dataset. ADPAc also provides binary datasets for these labels. These binary datasets include both positive and negative samples; each positive sample corresponds to the presence of a specific privacy behavior whereas a negative sample corresponds to its absence. For example, in the binary dataset of \texttt{Collecting} label, a positive code sample implements collecting behavior whereas a negative code sample does not. Each code sample is represented as an Abstract Syntax Tree (AST) containing paths within, where a path is a traversal of nodes in an AST. AST and its paths are explained in more detail in Appendix~\ref{subsec:appendix-ast-paths}. 

Additionally, as mentioned earlier, each code sample can include up to three hops. Therefore, for each label, there are three versions of the binary dataset, which include the same samples containing different numbers of hops. For example, \texttt{Collecting\_1\_Hop}, \texttt{Collecting\_2\_Hop}, and \texttt{Collecting\_3\_Hop} have the same samples but contain one (the first hop), two (the first two hops), and three (all the three) hops, respectively. Since there are eight \textit{Privacy Action} labels in PAcT and each has three versions, there are 24 datasets in total. In this work, we use these 24 binary datasets for our experiments and for training our multi-head encoder model. 




\subsection{Attention}
\label{subsec:attention}

The concept of attention was introduced by Bahdanau et al. \cite{bahdanau2014neural} as a method to jointly align and translate sequences which significantly improves the translation tasks. Subsequently, attention-based networks were used in several other NLP tasks and achieved state-of-the-art performance \cite{yang2019end, liu2019fine, yoshioka2021bert}. The idea behind attention is to quantitatively identify tokens in the input sequence that are more relevant to the task than other tokens. There are different variations of attention \cite{vaswani2017attention}. In this work, we use self-attention (or intra-attention); a mechanism in which the weights of the input tokens are determined based on the importance of other tokens in the same sequence.

\label{sec:background}

\section{Approach}

\begin{figure*}
    \centering
    \begin{subfigure}[b]{0.65\textwidth}
       \fbox{\includegraphics[width=\textwidth]{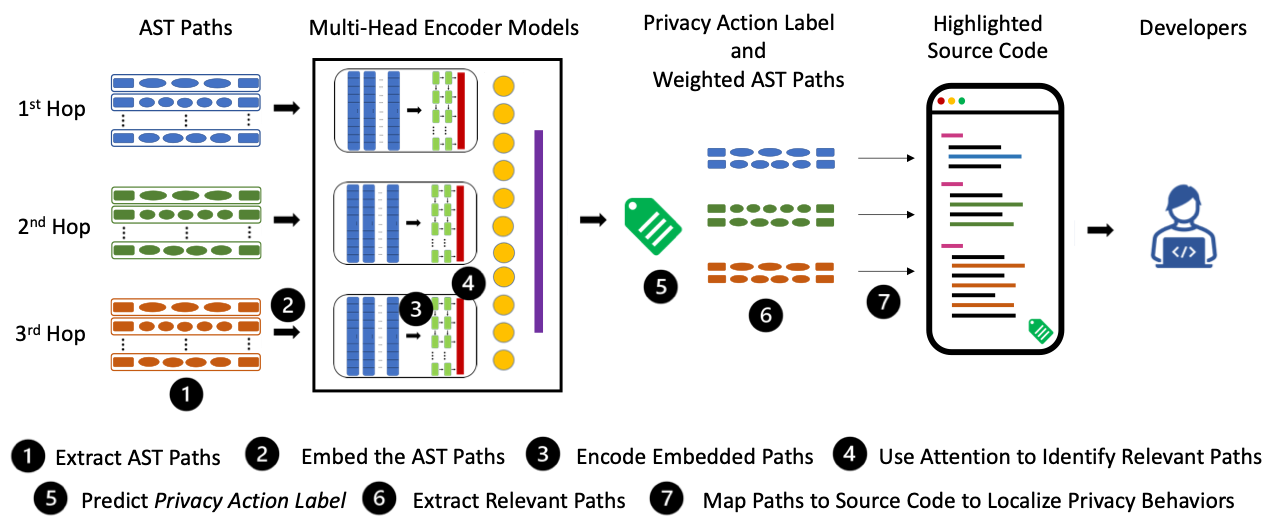}}
       \caption{An overview of our approach to classify and localize privacy behaviors.}
   \end{subfigure}
   \hfill
   \begin{subfigure}[b]{0.33\textwidth}
       \fbox{\includegraphics[width=\textwidth]{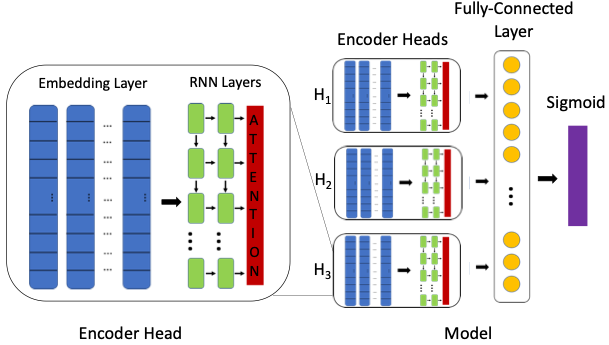}}
       \caption{Detailed architecture.}
   \end{subfigure}
   \hfill
   \caption{Overview of our approach and detailed architecture of multi-head encoder model}
      \label{fig:overview}
\end{figure*}

We now explain our approach to localizing privacy behaviors in application source code and describe the implementation details of our model. In summary, our approach works as follows: (i) extract AST paths for each hop in a code sample, (ii) embed the AST paths, (iii) encode embedded AST paths, (iv) use attention to identify relevant paths, (v) use fully connected layers to predict \textit{Privacy Action} label, (vi) extract relevant paths identified by the attention module, and (vii) map these paths to source code to finely localize privacy behaviors. Figure \ref{fig:overview} (a) shows an overview of our approach.

\subsection{Towards Fine-Grained Localization}
\label{subsec:model-overview}

We provide fine-grained localization of privacy behaviors by using attention weights to quantify relevant paths in each hop and then mapping these paths to source code. Towards this approach, the first step is to extract AST paths from each hop of a code sample. As described in Section \ref{subsec:pact-labels}, each code segment (PRCS) extracted from an application consists of three methods linked via a call graph. In line with other research studies that classify/summarize source code \cite{chen2019neural, alon2018code2seq}, we represent each code sample using AST paths. We extract these paths for each hop using a tool called ``\texttt{astminer}'' \cite{kovalenko2019pathminer}. 

The second and third steps are to embed and encode the extracted AST paths. Recall that each AST path consists of terminal and non-terminal nodes (as described in Appendix~\ref{subsec:appendix-ast-paths}). To embed each path, we first tokenize the paths into individual (terminal and non-terminal) nodes. We choose to tokenize non-terminal nodes based on our experimental results, as explained in Section \ref{subsec:tokenization-results}. Next, we embed each node using a pre-trained embedding model that we created using Gensim \cite{gensim}. The combined embedding of each node in an AST path is then passed to recurrent layers for encoding. To encode each hop, we use separate encoder heads to ensure that we preserve the semantic differences between the three methods. Since there are three methods in each sample, there are three heads in the multi-head encoder model. These multiple encoders are an upgrade from the baseline approach that used a single encoder head to encode all three hops.

In the fourth step, we need to identify relevant paths that implement privacy behaviors. This is necessary to predict \textit{Privacy Action} labels and finely localize privacy behaviors in the subsequent steps. To identify these relevant paths, we use attention. In each encoder head, an attention module uses a weighting mechanism that provides attention weights for each path in each hop to quantify the relevance of each encoded path.
This relevance indicates if a path implements a privacy behavior or not, i.e., paths with higher attention weights most likely use personal information which may help predict a label. 


In the fifth step, we predict \textit{Privacy Action} labels. We combine the attention weights with their respective encoded paths and then pass them to the fully-connected layers. This step creates a weighted representation of each \textit{hop} and provides the fully-connected layers to ``comprehend'' the privacy behaviors implemented across three hops. Using non-linearity, the output of the layers is converted into a prediction for a label. In the second last step towards fine-grained localization, we extract the output of the attention module (i.e., the attention weights) from each head and match them with their corresponding paths in each hop. This gives us the quantified relevance of each \textit{AST path} in each hop. We then sort these paths based on their weights and select 20 paths with the highest weights from each hop. We select this number based on experimental results that we explain in Section \ref{subsec:localization-results}. 

Lastly, we map these 20 paths from each hop to the source code using an automated script as follows: the script inspects the terminal nodes of each path, which contain the name of the identifier, and maps it to the line in the source code that contains the identifier. For example, in Appendix \ref{subsec:appendix-ast-paths} - Figure \ref{fig:ast} (c), the first AST path has a terminal node ``\texttt{getLastKnownLocation}''. Since the corresponding source code (Figure \ref{fig:ast} (a)) has the method name \texttt{getLastKnownLocation}, the first AST path is mapped to the highlighted statement in Figure \ref{fig:ast} (a). 

These mapped statements implement privacy behaviors and are highlighted in the code, thereby finely localizing privacy behaviors. These seven steps together provide an \emph{automated} fine-grained localization mechanism. 

We design our localization approach to identify privacy-relevant snippets at statement-level granularity since it provides maximum precision. A larger granularity, say a block of code comprised of several lines, may work in some cases where privacy-relevant statements are written together. In cases where privacy-relevant statements are spread out, this block-level granularity will highlight several non-relevant statements and result in a less precise localization. A statement-level granularity, thus, is a better approach since it identifies privacy-relevant code that are either written together or they are spread out.

\subsection{Implementation Details}
\label{subsec:model-architecture}

We now explain the implementation details of our steps: After we extract the AST paths for a code sample $C$, we randomly select $num\_paths$ paths. We decide how many paths to select from each hop based on our experiments (see Section \ref{subsec:num-paths-results}). Next, we tokenize the nodes in AST paths. Each AST path $p_i$ can be represented as $p_i = [t_s ; t_N ; t_e]$, where $t_s$ and $t_e$ are the start and terminal nodes, and $t_N = [t_1; ... ; t_8 ]$ is a list of non-terminal nodes. We found that in the ADPAc dataset, each sequence of non-terminal nodes can be of max length 8. Therefore, after tokenizing, each path $p_i$ can be represented as a list of 10 terminal and non-terminal nodes, $p_i = [t_s ; t_1; ... ; t_e]$. In case the length of non-terminal nodes is less than 8, we pad the AST path with zeros. We also pad each path with a zero to denote the end of the path. Hence, each tokenized path $p_i$ is 11 tokens long (the last token is padding). Since, each hop $h_i$ is represented by $num\_paths$ AST paths, $h_i = [p_1, p_2, ...p_{num\_paths}]$, where $i  \epsilon  [1, 2, 3]$. Mathematically, each hop $h_i$ is a $num\_paths \times 11$ matrix. Each code segment $C$ contains three hops, i.e., $C = [h_1; h_2; h_3]$, which makes it a  $3 \times num\_paths \times 11$ tensor.

As shown in Figure \ref{fig:overview} (b), each encoder head ($H_1$, $H_2$, and $H_3$) contains an embedding layer $E$, two recurrent layers, and an attention module. After tokenizing each path, we encode them: each individual hop $h_i$ in $C$ is passed to an encoder head $H_i$. Inside each head, each hop is embedded as $E_{h_{i}} = [E_{p_1}; E_{p_2} ... E_{p_{num\_paths}}]$, where $E_{h_{i}}$ is a $num\_paths \times 11 \times embed\_size$ tensor containing embedded AST paths. Each embedded path $E_{p_{i}}$ is represented as $E_{p_{i}} = [E_{t_s}; E_{t_1}; ...; E_{t_e}]$, where $E_{p_i}$ is a $11 \times embed\_size$ matrix and $E_{t_x}$ is the embedding of an individual node in an AST path which is an $embed\_size$ dimensional vector.

After embedding, we pass $E_{h_i}$ through the recurrent layers for encoding and use the output $L_i$ as the context and $hidden_i$ as the hidden state i.e., $L_i, hidden_i = RNN(E_{h_i})$. Here $L_i$ is $num\_paths \times embed\_size$ matrix and the hidden state is a $num\_paths$ size vector. We pass the hidden states to the attention module which returns attention weights of size $num\_paths$. We stack the attention weights from the three heads which give us a $3 \times num\_paths$ matrix. We pass this matrix through fully-connected layers and apply sigmoid non-linearity to get the classification probability. Lastly, we match the attention weights from each hop to the AST paths and use a script to map them to their source code.

\label{sec:approach}

\section{Experiments}

In this section, we describe the rationale for our experiments, their setup, and evaluation techniques.
 
\subsection{Research Questions}
\label{subsec:rq}

Our research objective is threefold: first, to find the optimal model configurations for classifying privacy behaviors in code snippets. Second, evaluate our model's performance in classifying privacy behaviors in comparison with other models. Third, evaluate the feasibility of our approach to finely localize privacy behaviors and its efficacy in helping to write privacy statements. For these objectives, we ask the following three research questions:

\textbf{RQ 1:} Which configurations provide the optimal performance for classification of \textit{Privacy Action} labels?

Based on the specific configuration, we ask the following sub-research questions: 

\textit{RQ 1.1: } How does the tokenization of non-terminal nodes affect the classification performance of the model?

\textit{RQ 1.2: } Which type of recurrent layers perform better? LSTM or Bi-LSTM?

\textit{RQ 1.3: } What is the optimal number of AST paths to represent each code sample?

\textbf{RQ 2}: Does the multi-head encoder model provide any quantitative increase in the classification performance as compared to other models?

\textbf{RQ 3}: Is our fine-grained localization efficacious?

We evaluate the efficacy of our approach to localize privacy behaviors and help write privacy statements qualitatively and quantitatively based on the following sub-research questions:

\textit{RQ 3.1} How does fine-grained localization help developers write privacy statements? 

\textit{RQ 3.2} What is the accuracy of fine-grained localization in identifying privacy behaviors?

The rationale for RQ 1 is based on the following two reasons: first, the baseline model \cite{jain2022pact} did not include many common model configurations that could improve the model's classification performance. Second, there is no systematic study that compares how the \textit{combination} of each of these components affects the model's performance for code classification. Therefore, we experiment with tokenization, Bi-LSTM layers, and the number of AST paths. We choose these configuration options for the following reasons. \textit{Tokenization (RQ 1.1):} In the baseline approach, AST paths are tokenized into three tokens where the terminal nodes are considered as two tokens and the non-terminal nodes are considered as a single token (see Appendix~\ref{subsec:appendix-ast-paths} for reference). By not tokenizing non-terminal nodes, the syntactic structural details of each code sample are diminished which could potentially hinder the classification accuracy. \textit{Bi-LSTM Layers (RQ 1.2):} The primary difference between LSTM and Bi-LSTM layers is the additional backward direction encoding of input sequences in Bi-LSTM layers which provides context surrounding each token in the sequence. Since an AST path can be traversed in either direction, including a reversed encoding may improve the classification accuracy. \textit{AST Paths (RQ 1.3)}: The number of paths used to represent a code sample is varied from 100 - 300 in code summarization studies \cite{haque2020improved, hu2018summarizing}. Therefore, to evaluate the optimal number of AST paths, especially for classification, we compare the performance of 100, 200, and 300 AST paths. It is possible to experiment with several other configuration choices, such as source code tokens versus AST paths, to learn their effects on a classification task; however, these are tangential to the goal of this paper since we focus on the feasibility of fine-grained localization.

In RQ 2, we aim to quantitatively evaluate whether our novel multi-head encoder model improves the performance of classifying \textit{Privacy Action} labels in comparison to the baseline model and its derivatives with modified configurations (i.e., models from RQ 1). 

In RQ 3, we evaluate fine-grained localization using qualitative (RQ 3.1) and quantitative (RQ 3.2) approaches. Qualitatively, we first analyze automated localization and then evaluate its dis/advantages in helping software professionals write privacy statements. Quantitatively, we evaluate the accuracy of our approach in identifying privacy behaviors. 

\subsection{Experimental Setup}
\label{subsec:experimental_setup}

We use the 24 binary datasets from ADPAc \cite{jain2022pact} to answer RQ 1 (\textit{Exp 1-6}). Recall that each of the 8 \textit{Privacy Action} labels contains 3 versions of the dataset (for example, \texttt{Collecting\_1\_Hop}, \texttt{Collecting\_2\_Hop}, and \texttt{Collecting\_3\_Hop}), with each version containing paths from one more hop than the previous version (see Section~\ref{subsec:pact-labels}). To answer RQs 1.1-1.3, we begin with the baseline model \cite{jain2022pact} and add attention (i.e.,\textit{ Exp 1: L\_100}). Based on the experiments, we make further configuration changes to this attention-based model. We intentionally add attention to the baseline model, since it is key to our approach and several studies demonstrate the efficacy of attention in improving the models' performance across different NLP tasks \cite{bahdanau2014neural, vaswani2017attention}. 

To limit the number of explored configurations that help us answer our RQs, we make incremental changes to determine the impact of each configuration and identify optimal selection rather than trying out all possible configuration combinations. We modify one configuration in each experiment, and based on the results, we keep this configuration fixed for subsequent experiments. For example, based on the results for RQ 1.1 (i.e., \textit{Exp 1: L\_100}), we make a decision to include/not-include the tokenization of non-terminal nodes. Then, when we evaluate RQ 1.2 and RQ 1.3, we use that configuration and only vary the choice of recurrent layer and the number of AST paths. 

To set up the experiments for RQ 1.1, (\textit{Exp 1: L\_100}), we tokenize each node in an AST path, including non-terminal nodes. For RQ 1.2, (\textit{Exp 4: Bi\_100}), we replace LSTM layers with Bi-LSTM in the model. For RQ 1.3 (\textit{Exp 1-6}), we randomly pool $N$ AST paths from each code sample for each version of the dataset. For example, if $N = 100$, for \texttt{Collecting\_1\_Hop} dataset, which contains AST paths from only the first hop, we randomly select 100 AST paths from each sample. For \texttt{Collecting\_2\_Hop} dataset, we pool 100 AST paths from the combined paths of the first two hops, and similarly, we pool 100 paths from the combined paths of all three hops in \texttt{Collecting\_3\_Hop} case. In these experiments, $N$ varies between 100 - 300 at 100-step increments. In case a sample contains less than $N$ paths, we pad the sample with zeros, which we call null paths. We summarize these various experiment configurations explored in  RQ 1 (and its sub-questions) in Table~\ref{table:experiment-configurations}. 

\begin{table}[t]
\centering
\caption{RQ 1 and RQ 2 experiment configurations. ``Tok'': tokenizing non-terminal nodes. ``Attn'': using attention. }
\begin{tabular}{|l|l|l|l|l|}
\hline
                      & \textbf{Tok} & \textbf{Attn} & \textbf{RNN Type} & \textbf{Paths} \\ \hline
\textbf{Baseline}     & False                    & False              & LSTM         & 100            \\ \hline
\textbf{Exp 1: L\_100} & True                     & True               & LSTM         & 100            \\ \hline
\textbf{Exp 2: L\_200} & True                     & True               & LSTM         & 200            \\   \hline
\textbf{Exp 3: L\_300} & True                     & True               & LSTM         & 300            \\ \hline
\textbf{Exp 4: Bi\_100} & True                     & True               & Bi-LSTM      & 100            \\ \hline
\textbf{Exp 5: Bi\_200} & True                     & True               & Bi-LSTM      & 200            \\ \hline
\textbf{Exp 6: Bi\_300} & True                     & True               & Bi-LSTM      & 300            \\ \hline
\textbf{Multi-Head Encoder} & True                     & True               & LSTM      & 100/hop            \\ \hline
\end{tabular}
\label{table:experiment-configurations}
\end{table}

To answer RQ 2 (\textit{Multi-Head Encoder}), we train our multi-head encoder model and compare its performance with the baseline model (i.e., Jain et al.~\cite{jain2022pact}) as well as the best configurations derived from RQ 1 (Table \ref{table:experiment-configurations}). Note that the RNN type for the multi-head encoder model in Table \ref{table:experiment-configurations} applies to each encoder head. To prepare the dataset, we extract the AST paths from 1\_Hop, 2\_Hop, and 3\_Hop datasets for each \textit{Privacy Action} label and separate the paths for each hop. For example, from \texttt{Collecting\_1\_Hop}, \texttt{Collecting\_2\_Hop}, and \texttt{Collecting\_3\_Hop} datasets, we separate the paths belonging only to the first hop, second hop, and third hop, respectively. Since this model is trained using paths from all three hops, we only compare the results with the 3\_Hop versions of the baseline and RQ 1 experiments (i.e., \textit{Exp 3 and Exp 6}).

We trained the models for RQ 1 and 2 using the following hyperparameters: a batch size of 8 and the Adam optimizer to modify the weights. The learning rate was fixed at 1e-5 and we used binary cross entropy to penalize the model. For each dataset, training, validation, and test sets were split in an 80:10:10 ratio. We chose these hyperparameters since these were also used in the baseline model \cite{jain2022pact}. For RQ 1, each model was trained for 50 epochs since we did not find improvement in the results after 50. For RQ 2, we varied the number of epochs since the model convergence differed for each label. For both RQs, after each epoch, the models were evaluated on the validation set. If they achieved better accuracy than the previous best epoch then the parameters were saved. We also monitored the training and validation accuracy to ensure the models did not overfit. After the training, the parameters from the best epoch were used on the test set for evaluation. We answer RQ 1 and 2 based on quantitative metrics, accuracy and F-1 scores. These are standard metrics used in classification tasks to compare the performance of different ML models. We balance our comparison by discussing overall patterns in classification accuracy for each configuration change while also highlighting interesting changes in the results of individual labels. To develop the model, we use PyTorch 1.8.1 with Python 3.7 and run all experiments on a workstation with a Xeon CPU, 54 GB RAM, and a Tesla T4 GPU. 

To answer RQ 3, we first conduct a manual inspection of mappings of some samples to analyze and draw insights into our approach to finely localize privacy behaviors. Next, to answer the two sub-research questions, we select 20 random samples from the test set of all labels, use our script to localize privacy behaviors in them, i.e., map the AST paths with the highest attention weights to the source code, and highlight them. 


To qualitatively evaluate fine-grained localization in RQ3.1, we asked six software professionals at our university to write privacy statements for each sample, explaining \textit{how} the personal information is being used and \textit{why}. To investigate any differences in the statements written when the samples are localized versus when they are not, we divided the six annotators (i.e., software professionals), into two groups based on their experience. Annotators in Group \#1 (i.e., Annotators \#1, \#2, and \#3), have more than four years of software development experience and two years of experience in privacy research. Whereas Group \#2 annotators have about two years of software development experience and approximately one year of experience in privacy research. Each group was given 20 samples, of which only 10 were finely localized. Moreover, the samples that were localized for Group \#1 were not localized for Group \#2, resulting in privacy statements for the same sample that were written with and without localization. 

Apart from writing privacy statements, we also asked each annotator if and how the localization helped, and the time it took them to write each statement. We compare privacy statements and the time taken between localized and non-localized samples of similar length for each annotator as well as between the two groups for the same samples. These comparisons help us understand how fine-grained localization affects annotators individually, as well as among different populations (i.e., professionals with little privacy experience vs. those with more experience).

In RQ 3.2, we evaluated the accuracy of fine-grained localization by using the same 20 samples from RQ 3.1, all of which were finely localized. We then asked annotators in Group \#1 who are privacy experts to evaluate if the highlighted statements implement privacy behaviors by providing a binary response (`yes'/`no') for each highlighted statement. Since each sample consists of three methods (see Section \ref{sec:background}), we evaluated a total of 60 methods where 230 of their statements were highlighted. To measure the inter-rater agreement among the annotators we used Krripendorff's Alpha\footnote{\url{https://www.statisticshowto.com/krippendorffs-alpha/}} and Fleiss's Kappa\footnote{\url{https://www.statisticshowto.com/fleiss-kappa/}}.

\label{sec:experiments}

\section{Results}

In this section, we report our results and answer our research questions. Table \ref{table:config-experiments-results} shows the quantitative results for RQ~1 and RQ 2, which includes the baseline results \cite{jain2022pact} in the first column (\textit{Baseline}). We show the confusion matrices for RQ~2 in Appendix \ref{subsec:appendix-rq-1.2} - Figures  \ref{fig:result-cm-practice} and \ref{fig:result-cm-purpose}. 
For RQ~3, we first discuss automated fine-grained localization with one representative code sample shown in Figure \ref{fig:result-src-code}, and its weighted AST paths in Figure \ref{fig:result-ast-paths}. We, then, discuss dis/advantages of our approach in helping software professionals write privacy statements with a representative sample in Fig.~\ref{fig:result-rq-3.1-long-good} (additional examples are in Figure~\ref{fig:result-rq-3.1-small-bad}-\ref{fig:result-rq-3.1-long-bad} in Appendix~\ref{subsec:appendix-rq-3.1}). Lastly, we report the accuracy of our approach for RQ 3.2 (examples shown in Figures \ref{fig:result-rq3-1} and \ref{fig:result-rq3-2} in Appendix~\ref{subsec:appendix-rq-3.2}). 

\subsection{RQ 1: Optimal Configurations}

As stated in Section \ref{sec:experiments}, to answer sub-research questions, we make incremental changes to the configuration starting with the baseline model. Overall, we find an attention-based model with tokenization of non-terminal nodes helps significantly improve the performance. 
Between LSTM and Bi-LSTM layers, the difference is insignificant when we use fewer paths (say 100), but as we increase this number (to say, 300), Bi-LSTM provides better results. 300 paths are an optimal choice to represent each code sample, and 100 paths to represent each hop. 

\subsubsection{RQ 1.1: Tokenization of Non-Terminal Nodes}
\label{subsec:tokenization-results}

The results for RQ 1.1 (\textit{Exp 1:L\_100} column in Table \ref{table:config-experiments-results}) indicate that by tokenizing non-terminal nodes (and using attention), we noticeably increase the accuracy for most labels. We observe that accuracy increases by $\sim$5\% on average for \textit{Practice} and \textit{Purpose} labels with the F-1 score also increasing for most labels. For some labels, the increase in accuracy is more significant; for example with \texttt{Processing\_1\_Hop} and \texttt{Advertisement\_1\_Hop}, the accuracy increases by 12\% and 15\%, respectively. These improvements in the scores can be attributed to the increase in the syntactic structural information of a code sample, which is what non-terminal nodes represent. 

For example, a code sample `\texttt{Processing}' personal information, say location, executes several operations ranging from \textit{comparing} accuracy of location providers to \textit{calculating} distance to the user's address. The syntactic structural information of these operations is captured by the tokenized non-terminal nodes, and when provided to the model, it can help predict \texttt{Processing} with better accuracy. Additionally, attention helps the model learn which structures contribute to \texttt{Processing} and which ones do not. To test the efficacy of attention, we ran an experiment in which we tokenized non-terminal nodes and trained using the baseline model (i.e., without attention) \cite{jain2022pact}). In this experiment, we noticed an average accuracy of 59.85\% and 54.40\% for \textit{Practice} and \textit{Purpose} labels, which are 7\% and 20\% lower than the baseline models (i.e., \textit{Baseline} column in Table \ref{table:config-experiments-results}). This noticeable decrease in the performance indicates the significance of attention, especially, with tokenized non-terminal nodes. 

Interestingly, we found a noticeable decrease in F-1 scores for \texttt{Sharing} in \textit{Exp 1:L\_100}. This decrease is most likely due to an information overload of syntactic structures. Since `\texttt{Sharing}' often occurs with calls to third-party libraries or third-party libraries calling permission-requiring APIs~\cite{jain2022pact}, such information is embedded in the identifiers of source code, i.e., the terminal nodes of AST paths.

\begin{landscape}
\begin{table}[]
\caption{Accuracy and F-1 Scores for each experiment conducted for RQs 1 and 2. All experiments except the baseline, tokenize non-terminal nodes and use attention. We use ``L'' to denote use of LSTM layers, ``Bi'' for Bi-LSTM layers, and ``100|200|300'' are the number of AST paths used in the experiment. See Table \ref{table:experiment-configurations} for details. The color of the cell denotes the range of the score. {\color[HTML]{FD6864}Red}: 50-60\%, {\color[HTML]{FFCB2F}Orange}: 60-70\%, {\color[HTML]{EBFF0C}Lime}: 70-80\%, {\color[HTML]{67FD9A}Light Green}: 80-90\%, {\color [HTML]{32CB00}Green}: 90+\%. A row at the end of each category shows the average score for all datasets in that category. }
\resizebox{\columnwidth}{!}{%
\begin{tabular}{|l|ll|ll|ll|ll|ll|ll|ll|ll|}
\hline
\multicolumn{1}{|l|}{Dataset}                & \multicolumn{2}{c|}{Baseline}                                                                               & \multicolumn{2}{c|}{Exp 1: L\_100}                                                                & \multicolumn{2}{c|}{Exp 2: L\_200}                                                               & \multicolumn{2}{c|}{Exp 3: L\_300}                                                               & \multicolumn{2}{c|}{Exp 4: Bi\_100}                                                              & \multicolumn{2}{c|}{Exp 5: Bi\_200}                                                              & \multicolumn{2}{c|}{Exp 6: Bi\_300}                                                              & \multicolumn{2}{l|}{Multi-Head Encoder}                                                      \\ \hline
\multicolumn{1}{|l|}{}                       & \multicolumn{1}{l|}{Acc}                             & \multicolumn{1}{l|}{F1}                              & \multicolumn{1}{l|}{Acc}                             & \multicolumn{1}{l|}{F1}                              & \multicolumn{1}{l|}{Acc}                             & \multicolumn{1}{l|}{F1}                              & \multicolumn{1}{l|}{Acc}                             & \multicolumn{1}{l|}{F1}                              & \multicolumn{1}{l|}{Acc}                             & \multicolumn{1}{l|}{F1}                              & \multicolumn{1}{l|}{Acc}                             & \multicolumn{1}{l|}{F1}                              & \multicolumn{1}{l|}{Acc}                             & \multicolumn{1}{l|}{F1}                              & \multicolumn{1}{l|}{Acc}                             & F1                              \\ \hline
\multicolumn{17}{|c|}{\textit{Practice}}                                                                                                                                                                                                                                                                                                                                                                                                                                                                                                                                                                                                                                                                                                                                                                                                                                                                        \\ \hline
\textit{Collecting}    & \multicolumn{1}{l|}{}                                &                                 & \multicolumn{1}{l|}{}                                &                                 & \multicolumn{1}{l|}{}                                &                                 & \multicolumn{1}{l|}{}                                &                                 & \multicolumn{1}{l|}{}                                &                                 & \multicolumn{1}{l|}{}                                &                                 & \multicolumn{1}{l|}{}                                &                                 & \multicolumn{1}{l|}{}                                &                                 \\ \hline
1 Hop                  & \multicolumn{1}{l|}{\cellcolor[HTML]{EBFF0C}70.39\%} & \cellcolor[HTML]{EBFF0C}74.58\% & \multicolumn{1}{l|}{\cellcolor[HTML]{EBFF0C}75.00\%} & \cellcolor[HTML]{EBFF0C}72.46\% & \multicolumn{1}{l|}{\cellcolor[HTML]{EBFF0C}73.03\%} & \cellcolor[HTML]{EBFF0C}70.07\% & \multicolumn{1}{l|}{\cellcolor[HTML]{EBFF0C}72.37\%} & \cellcolor[HTML]{FFCB2F}69.12\% & \multicolumn{1}{l|}{\cellcolor[HTML]{EBFF0C}73.68\%} & \cellcolor[HTML]{EBFF0C}71.01\% & \multicolumn{1}{l|}{\cellcolor[HTML]{EBFF0C}74.34\%} & \cellcolor[HTML]{EBFF0C}71.94\% & \multicolumn{1}{l|}{\cellcolor[HTML]{EBFF0C}72.37\%} & \cellcolor[HTML]{FFCB2F}69.12\% & \multicolumn{1}{l|}{-}                               & -                               \\
2 Hop                  & \multicolumn{1}{l|}{\cellcolor[HTML]{FFCB2F}68.42\%} & \cellcolor[HTML]{EBFF0C}73.03\% & \multicolumn{1}{l|}{\cellcolor[HTML]{EBFF0C}74.34\%} & \cellcolor[HTML]{EBFF0C}71.53\% & \multicolumn{1}{l|}{\cellcolor[HTML]{EBFF0C}75.66\%} & \cellcolor[HTML]{EBFF0C}73.38\% & \multicolumn{1}{l|}{\cellcolor[HTML]{EBFF0C}75.66\%} & \cellcolor[HTML]{EBFF0C}74.48\% & \multicolumn{1}{l|}{\cellcolor[HTML]{EBFF0C}75.66\%} & \cellcolor[HTML]{EBFF0C}73.38\% & \multicolumn{1}{l|}{\cellcolor[HTML]{EBFF0C}76.97\%} & \cellcolor[HTML]{EBFF0C}74.82\% & \multicolumn{1}{l|}{\cellcolor[HTML]{67FD9A}84.87\%} & \cellcolor[HTML]{32CB00}91.81\% & \multicolumn{1}{l|}{-}                               & -                               \\
3 Hop                  & \multicolumn{1}{l|}{\cellcolor[HTML]{EBFF0C}71.09\%} & \cellcolor[HTML]{EBFF0C}74.36\% & \multicolumn{1}{l|}{\cellcolor[HTML]{EBFF0C}76.32\%} & \cellcolor[HTML]{EBFF0C}73.53\% & \multicolumn{1}{l|}{\cellcolor[HTML]{EBFF0C}76.32\%} & \cellcolor[HTML]{EBFF0C}73.53\% & \multicolumn{1}{l|}{\cellcolor[HTML]{EBFF0C}75.66\%} & \cellcolor[HTML]{EBFF0C}73.38\% & \multicolumn{1}{l|}{\cellcolor[HTML]{EBFF0C}76.32\%} & \cellcolor[HTML]{EBFF0C}73.13\% & \multicolumn{1}{l|}{\cellcolor[HTML]{EBFF0C}77.63\%} & \cellcolor[HTML]{EBFF0C}73.85\% & \multicolumn{1}{l|}{\cellcolor[HTML]{67FD9A}84.87\%} & \cellcolor[HTML]{32CB00}91.81\% & \multicolumn{1}{c|}{\cellcolor[HTML]{32CB00}98.03\%} & \cellcolor[HTML]{32CB00}98.04\% \\ \hline

\textit{Sharing}       & \multicolumn{1}{l|}{}                                &                                 & \multicolumn{1}{l|}{}                                &                                 & \multicolumn{1}{l|}{}                                &                                 & \multicolumn{1}{l|}{}                                &                                 & \multicolumn{1}{l|}{}                                &                                 & \multicolumn{1}{l|}{}                                &                                 & \multicolumn{1}{l|}{}                                &                                 & \multicolumn{1}{l|}{}                                &                                 \\ \hline
1 Hop                  & \multicolumn{1}{l|}{\cellcolor[HTML]{FFCB2F}65.10\%} & \cellcolor[HTML]{FFCB2F}69.82\% & \multicolumn{1}{l|}{\cellcolor[HTML]{FFCB2F}66.41\%} & \cellcolor[HTML]{FD6864}54.74\% & \multicolumn{1}{l|}{\cellcolor[HTML]{FFCB2F}67.97\%} & \cellcolor[HTML]{FD6864}53.23\% & \multicolumn{1}{l|}{\cellcolor[HTML]{FFCB2F}67.71\%} & \cellcolor[HTML]{FD6864}56.64\% & \multicolumn{1}{l|}{\cellcolor[HTML]{FFCB2F}66.67\%} & \cellcolor[HTML]{FD6864}52.59\% & \multicolumn{1}{l|}{\cellcolor[HTML]{FFCB2F}66.67\%} & \cellcolor[HTML]{FD6864}55.24\% & \multicolumn{1}{l|}{\cellcolor[HTML]{FFCB2F}63.02\%} & \cellcolor[HTML]{EBFF0C}77.10\% & \multicolumn{1}{l|}{-}                               & -                               \\
2 Hop                  & \multicolumn{1}{l|}{\cellcolor[HTML]{FFCB2F}67.10\%} & \cellcolor[HTML]{EBFF0C}73.39\% & \multicolumn{1}{l|}{\cellcolor[HTML]{FFCB2F}66.41\%} & \cellcolor[HTML]{FD6864}52.40\% & \multicolumn{1}{l|}{\cellcolor[HTML]{FFCB2F}69.27\%} & \cellcolor[HTML]{FD6864}57.55\% & \multicolumn{1}{l|}{\cellcolor[HTML]{FFCB2F}68.75\%} & \cellcolor[HTML]{FD6864}56.20\% & \multicolumn{1}{l|}{\cellcolor[HTML]{FFCB2F}66.93\%} & \cellcolor[HTML]{FD6864}51.71\% & \multicolumn{1}{l|}{\cellcolor[HTML]{FFCB2F}67.97\%} & \cellcolor[HTML]{FD6864}55.91\% & \multicolumn{1}{l|}{\cellcolor[HTML]{FFCB2F}64.58\%} & \cellcolor[HTML]{EBFF0C}75.89\% & \multicolumn{1}{l|}{-}                               & -                               \\
3 Hop                  & \multicolumn{1}{l|}{\cellcolor[HTML]{EBFF0C}71.09\%} & \cellcolor[HTML]{EBFF0C}77.58\% & \multicolumn{1}{l|}{\cellcolor[HTML]{FFCB2F}68.23\%} & \cellcolor[HTML]{FD6864}55.15\% & \multicolumn{1}{l|}{\cellcolor[HTML]{FFCB2F}67.71\%} & \cellcolor[HTML]{FD6864}55.71\% & \multicolumn{1}{l|}{\cellcolor[HTML]{FFCB2F}68.49\%} & \cellcolor[HTML]{FD6864}56.94\% & \multicolumn{1}{l|}{\cellcolor[HTML]{FFCB2F}68.49\%} & \cellcolor[HTML]{FD6864}55.68\% & \multicolumn{1}{l|}{\cellcolor[HTML]{FFCB2F}68.23\%} & \cellcolor[HTML]{FD6864}57.64\% & \multicolumn{1}{l|}{\cellcolor[HTML]{FFCC67}64.32\%} & \cellcolor[HTML]{EBFF0C}77.36\% & \multicolumn{1}{l|}{\cellcolor[HTML]{32CB00}91.41\%} & \cellcolor[HTML]{32CB00}90.32\% \\ \hline
\textit{Processing}    & \multicolumn{1}{l|}{}                                &                                 & \multicolumn{1}{l|}{}                                &                                 & \multicolumn{1}{l|}{}                                & \cellcolor[HTML]{FFFFFF}        & \multicolumn{1}{l|}{}                                & \cellcolor[HTML]{FFFFFF}        & \multicolumn{1}{l|}{}                                &                                 & \multicolumn{1}{l|}{}                                &                                 & \multicolumn{1}{l|}{}                                &                                 & \multicolumn{1}{l|}{}                                &                                 \\ \hline
1 Hop                  & \multicolumn{1}{l|}{\cellcolor[HTML]{FD6864}58.75\%} & \cellcolor[HTML]{FD6864}51.04\% & \multicolumn{1}{l|}{\cellcolor[HTML]{EBFF0C}70.50\%} & \cellcolor[HTML]{EBFF0C}74.89\% & \multicolumn{1}{l|}{\cellcolor[HTML]{FFCB2F}69.75\%} & \cellcolor[HTML]{EBFF0C}71.93\% & \multicolumn{1}{l|}{\cellcolor[HTML]{EBFF0C}71.75\%} & \cellcolor[HTML]{EBFF0C}73.16\% & \multicolumn{1}{l|}{\cellcolor[HTML]{EBFF0C}71.00\%} & \cellcolor[HTML]{EBFF0C}75.93\% & \multicolumn{1}{l|}{\cellcolor[HTML]{EBFF0C}71.50\%} & \cellcolor[HTML]{EBFF0C}72.73\% & \multicolumn{1}{l|}{\cellcolor[HTML]{EBFF0C}71.50\%} & \cellcolor[HTML]{EBFF0C}79.20\% & \multicolumn{1}{l|}{-}                               & -                               \\
2 Hop                  & \multicolumn{1}{l|}{\cellcolor[HTML]{FFCB2F}64.75\%} & \cellcolor[HTML]{FD6864}59.83\% & \multicolumn{1}{l|}{\cellcolor[HTML]{EBFF0C}72.50\%} & \cellcolor[HTML]{EBFF0C}76.89\% & \multicolumn{1}{l|}{\cellcolor[HTML]{EBFF0C}71.00\%} & \cellcolor[HTML]{EBFF0C}77.17\% & \multicolumn{1}{l|}{\cellcolor[HTML]{EBFF0C}72.25\%} & \cellcolor[HTML]{EBFF0C}76.03\% & \multicolumn{1}{l|}{\cellcolor[HTML]{EBFF0C}71.50\%} & \cellcolor[HTML]{EBFF0C}76.25\% & \multicolumn{1}{l|}{\cellcolor[HTML]{EBFF0C}72.25\%} & \cellcolor[HTML]{EBFF0C}76.43\% & \multicolumn{1}{l|}{\cellcolor[HTML]{EBFF0C}73.50\%} & \cellcolor[HTML]{67FD9A}80.87\% & \multicolumn{1}{l|}{-}                               & -                               \\
3 Hop                  & \multicolumn{1}{l|}{\cellcolor[HTML]{FFCB2F}64.25\%} & \cellcolor[HTML]{FD6864}59.49\% & \multicolumn{1}{l|}{\cellcolor[HTML]{EBFF0C}72.25\%} & \cellcolor[HTML]{EBFF0C}77.02\% & \multicolumn{1}{l|}{\cellcolor[HTML]{EBFF0C}71.50\%} & \cellcolor[HTML]{EBFF0C}76.64\% & \multicolumn{1}{l|}{\cellcolor[HTML]{EBFF0C}73.50\%} & \cellcolor[HTML]{EBFF0C}75.80\% & \multicolumn{1}{l|}{\cellcolor[HTML]{EBFF0C}72.00\%} & \cellcolor[HTML]{EBFF0C}76.95\% & \multicolumn{1}{l|}{\cellcolor[HTML]{EBFF0C}73.50\%} & \cellcolor[HTML]{EBFF0C}75.80\% & \multicolumn{1}{l|}{\cellcolor[HTML]{EBFF0C}74.00\%} & \cellcolor[HTML]{67FD9A}81.56\% & \multicolumn{1}{l|}{\cellcolor[HTML]{32CB00}94.25\%} & \cellcolor[HTML]{32CB00}94.56\% \\ \hline
\textit{Others}        & \multicolumn{1}{l|}{}                                &                                 & \multicolumn{1}{l|}{}                                &                                 & \multicolumn{1}{l|}{}                                &                                 & \multicolumn{1}{l|}{}                                &                                 & \multicolumn{1}{l|}{}                                &                                 & \multicolumn{1}{l|}{}                                &                                 & \multicolumn{1}{l|}{}                                &                                 & \multicolumn{1}{l|}{}                                &                                 \\ \hline
1 Hop                  & \multicolumn{1}{l|}{\cellcolor[HTML]{FFCB2F}66.50\%} & \cellcolor[HTML]{FD6864}59.06\% & \multicolumn{1}{l|}{\cellcolor[HTML]{EBFF0C}71.71\%} & \cellcolor[HTML]{EBFF0C}71.52\% & \multicolumn{1}{l|}{\cellcolor[HTML]{EBFF0C}71.05\%} & \cellcolor[HTML]{EBFF0C}70.67\% & \multicolumn{1}{l|}{\cellcolor[HTML]{EBFF0C}73.03\%} & \cellcolor[HTML]{EBFF0C}72.11\% & \multicolumn{1}{l|}{\cellcolor[HTML]{EBFF0C}70.39\%} & \cellcolor[HTML]{FFCB2F}69.39\% & \multicolumn{1}{l|}{\cellcolor[HTML]{EBFF0C}71.71\%} & \cellcolor[HTML]{EBFF0C}70.75\% & \multicolumn{1}{l|}{\cellcolor[HTML]{67FD9A}88.16\%} & \cellcolor[HTML]{32CB00}93.62\% & \multicolumn{1}{l|}{-}                               & -                               \\
2 Hop                  & \multicolumn{1}{l|}{\cellcolor[HTML]{FFCB2F}69.91\%} & \cellcolor[HTML]{FFCB2F}68.87\% & \multicolumn{1}{l|}{\cellcolor[HTML]{EBFF0C}72.37\%} & \cellcolor[HTML]{EBFF0C}71.62\% & \multicolumn{1}{l|}{\cellcolor[HTML]{EBFF0C}71.71\%} & \cellcolor[HTML]{EBFF0C}72.26\% & \multicolumn{1}{l|}{\cellcolor[HTML]{EBFF0C}73.68\%} & \cellcolor[HTML]{EBFF0C}73.68\% & \multicolumn{1}{l|}{\cellcolor[HTML]{FFCB2F}69.08\%} & \cellcolor[HTML]{EBFF0C}70.44\% & \multicolumn{1}{l|}{\cellcolor[HTML]{EBFF0C}75.00\%} & \cellcolor[HTML]{EBFF0C}75.64\% & \multicolumn{1}{l|}{\cellcolor[HTML]{67FD9A}88.82\%} & \cellcolor[HTML]{32CB00}93.99\% & \multicolumn{1}{l|}{-}                               & -                               \\
3 Hop                  & \multicolumn{1}{l|}{\cellcolor[HTML]{FFCB2F}69.00\%} & \cellcolor[HTML]{EBFF0C}72.99\% & \multicolumn{1}{l|}{\cellcolor[HTML]{EBFF0C}75.00\%} & \cellcolor[HTML]{EBFF0C}75.00\% & \multicolumn{1}{l|}{\cellcolor[HTML]{EBFF0C}71.71\%} & \cellcolor[HTML]{EBFF0C}71.90\% & \multicolumn{1}{l|}{\cellcolor[HTML]{EBFF0C}74.34\%} & \cellcolor[HTML]{EBFF0C}73.83\% & \multicolumn{1}{l|}{\cellcolor[HTML]{EBFF0C}74.34\%} & \cellcolor[HTML]{EBFF0C}72.73\% & \multicolumn{1}{l|}{\cellcolor[HTML]{EBFF0C}73.68\%} & \cellcolor[HTML]{EBFF0C}72.60\% & \multicolumn{1}{l|}{\cellcolor[HTML]{F8FF00}74.34\%} & \cellcolor[HTML]{F8FF00}74.84\% & \multicolumn{1}{l|}{\cellcolor[HTML]{32CB00}94.08\%} & \cellcolor[HTML]{32CB00}94.67\% \\ \hline  \hline
\textbf{Avg.}          & \multicolumn{1}{l|}{\cellcolor[HTML]{FFCB2F}67.20\%} & \cellcolor[HTML]{FFCB2F}67.84\% & \multicolumn{1}{l|}{\cellcolor[HTML]{EBFF0C}71.75\%} & \cellcolor[HTML]{FFCB2F}68.90\% & \multicolumn{1}{l|}{\cellcolor[HTML]{EBFF0C}71.39\%} & \cellcolor[HTML]{FFCB2F}68.67\% & \multicolumn{1}{l|}{\cellcolor[HTML]{EBFF0C}72.27\%} & \cellcolor[HTML]{FFCB2F}69.28\% & \multicolumn{1}{l|}{\cellcolor[HTML]{EBFF0C}71.34\%} & \cellcolor[HTML]{FFCB2F}68.27\% & \multicolumn{1}{l|}{\cellcolor[HTML]{EBFF0C}72.45\%} & \cellcolor[HTML]{FFCB2F}69.45\% & \multicolumn{1}{l|}{\cellcolor[HTML]{EBFF0C}75.36\%} & \cellcolor[HTML]{67FD9A}82.26\% & \multicolumn{1}{l|}{\cellcolor[HTML]{32CB00}94.44\%} & \cellcolor[HTML]{32CB00}94.39\% \\ \hline  \hline
\multicolumn{17}{|c|}{\textit{Purpose}}                                                                                                                                                                                                                                                                                                                                                                                                                                                                                                                                                                                                                                                                                                                                                                                                                                                                         \\ \hline
\textit{Functionality} & \multicolumn{1}{l|}{}                                &                                 & \multicolumn{1}{l|}{}                                &                                 & \multicolumn{1}{l|}{}                                &                                 & \multicolumn{1}{l|}{}                                &                                 & \multicolumn{1}{l|}{}                                &                                 & \multicolumn{1}{l|}{}                                &                                 & \multicolumn{1}{l|}{}                                &                                 & \multicolumn{1}{l|}{}                                &                                 \\ \hline
1 Hop                  & \multicolumn{1}{l|}{\cellcolor[HTML]{67FD9A}85.90\%} & \cellcolor[HTML]{67FD9A}88.89\% & \multicolumn{1}{l|}{\cellcolor[HTML]{67FD9A}89.89\%} & \cellcolor[HTML]{32CB00}90.21\% & \multicolumn{1}{l|}{\cellcolor[HTML]{67FD9A}88.83\%} & \cellcolor[HTML]{67FD9A}89.18\% & \multicolumn{1}{l|}{\cellcolor[HTML]{67FD9A}89.63\%} & \cellcolor[HTML]{67FD9A}89.82\% & \multicolumn{1}{l|}{\cellcolor[HTML]{67FD9A}89.63\%} & \cellcolor[HTML]{67FD9A}89.87\% & \multicolumn{1}{l|}{\cellcolor[HTML]{67FD9A}89.89\%} & \cellcolor[HTML]{67FD9A}89.95\% & \multicolumn{1}{l|}{\cellcolor[HTML]{32CB00}90.69\%} & \cellcolor[HTML]{32CB00}90.57\% & \multicolumn{1}{l|}{-}                               & -                               \\
2 Hop                  & \multicolumn{1}{l|}{\cellcolor[HTML]{67FD9A}86.97\%} & \cellcolor[HTML]{67FD9A}87.59\% & \multicolumn{1}{l|}{\cellcolor[HTML]{67FD9A}89.36\%} & \cellcolor[HTML]{32CB00}90.05\% & \multicolumn{1}{l|}{\cellcolor[HTML]{67FD9A}88.83\%} & \cellcolor[HTML]{67FD9A}89.34\% & \multicolumn{1}{l|}{\cellcolor[HTML]{32CB00}91.76\%} & \cellcolor[HTML]{32CB00}91.91\% & \multicolumn{1}{l|}{\cellcolor[HTML]{67FD9A}88.30\%} & \cellcolor[HTML]{67FD9A}89.00\% & \multicolumn{1}{l|}{\cellcolor[HTML]{32CB00}92.02\%} & \cellcolor[HTML]{32CB00}92.19\% & \multicolumn{1}{l|}{\cellcolor[HTML]{32CB00}92.55\%} & \cellcolor[HTML]{32CB00}92.55\% & \multicolumn{1}{l|}{-}                               & -                               \\
3 Hop                  & \multicolumn{1}{l|}{\cellcolor[HTML]{67FD9A}85.64\%} & \cellcolor[HTML]{67FD9A}86.43\% & \multicolumn{1}{l|}{\cellcolor[HTML]{32CB00}90.43\%} & \cellcolor[HTML]{32CB00}90.77\% & \multicolumn{1}{l|}{\cellcolor[HTML]{32CB00}91.22\%} & \cellcolor[HTML]{32CB00}91.52\% & \multicolumn{1}{l|}{\cellcolor[HTML]{32CB00}90.69\%} & \cellcolor[HTML]{32CB00}90.86\% & \multicolumn{1}{l|}{\cellcolor[HTML]{32CB00}90.69\%} & \cellcolor[HTML]{32CB00}90.96\% & \multicolumn{1}{l|}{\cellcolor[HTML]{32CB00}92.29\%} & \cellcolor[HTML]{32CB00}92.54\% & \multicolumn{1}{l|}{\cellcolor[HTML]{32CB00}92.29\%} & \cellcolor[HTML]{32CB00}92.39\% & \multicolumn{1}{l|}{\cellcolor[HTML]{32CB00}96.81\%} & \cellcolor[HTML]{32CB00}96.83\% \\ \hline
\textit{Advertisement} & \multicolumn{1}{l|}{}                                &                                 & \multicolumn{1}{l|}{}                                &                                 & \multicolumn{1}{l|}{}                                &                                 & \multicolumn{1}{l|}{}                                &                                 & \multicolumn{1}{l|}{}                                &                                 & \multicolumn{1}{l|}{}                                &                                 & \multicolumn{1}{l|}{}                                &                                 & \multicolumn{1}{l|}{}                                &                                 \\ \hline
1 Hop                  & \multicolumn{1}{l|}{\cellcolor[HTML]{EBFF0C}70.24\%} & \cellcolor[HTML]{EBFF0C}73.96\% & \multicolumn{1}{l|}{\cellcolor[HTML]{67FD9A}84.52\%} & \cellcolor[HTML]{67FD9A}80.88\% & \multicolumn{1}{l|}{\cellcolor[HTML]{67FD9A}85.71\%} & \cellcolor[HTML]{67FD9A}82.61\% & \multicolumn{1}{l|}{\cellcolor[HTML]{67FD9A}86.90\%} & \cellcolor[HTML]{67FD9A}84.29\% & \multicolumn{1}{l|}{\cellcolor[HTML]{67FD9A}84.52\%} & \cellcolor[HTML]{67FD9A}80.88\% & \multicolumn{1}{l|}{\cellcolor[HTML]{67FD9A}86.31\%} & \cellcolor[HTML]{67FD9A}83.21\% & \multicolumn{1}{l|}{\cellcolor[HTML]{67FD9A}83.33\%} & \cellcolor[HTML]{32CB00}90.60\% & \multicolumn{1}{l|}{-}                               & -                               \\
2 Hop                  & \multicolumn{1}{l|}{\cellcolor[HTML]{EBFF0C}78.57\%} & \cellcolor[HTML]{67FD9A}80\%    & \multicolumn{1}{l|}{\cellcolor[HTML]{67FD9A}85\%}    & \cellcolor[HTML]{67FD9A}80.60\% & \multicolumn{1}{l|}{\cellcolor[HTML]{67FD9A}82.74\%} & \cellcolor[HTML]{EBFF0C}77.17\% & \multicolumn{1}{l|}{\cellcolor[HTML]{67FD9A}84.52\%} & \cellcolor[HTML]{67FD9A}80.60\% & \multicolumn{1}{l|}{\cellcolor[HTML]{67FD9A}83.33\%} & \cellcolor[HTML]{EBFF0C}78.46\% & \multicolumn{1}{l|}{\cellcolor[HTML]{67FD9A}83.93\%} & \cellcolor[HTML]{EBFF0C}79.39\% & \multicolumn{1}{l|}{\cellcolor[HTML]{67FD9A}82.74\%} & \cellcolor[HTML]{32CB00}90.24\% & \multicolumn{1}{l|}{-}                               & -                               \\
3 Hop                  & \multicolumn{1}{l|}{\cellcolor[HTML]{EBFF0C}79.76\%} & \cellcolor[HTML]{67FD9A}80.46\% & \multicolumn{1}{l|}{\cellcolor[HTML]{67FD9A}85.71\%} & \cellcolor[HTML]{67FD9A}82.09\% & \multicolumn{1}{l|}{\cellcolor[HTML]{67FD9A}83.93\%} & \cellcolor[HTML]{EBFF0C}79.07\% & \multicolumn{1}{l|}{\cellcolor[HTML]{67FD9A}86.90\%} & \cellcolor[HTML]{67FD9A}84.29\% & \multicolumn{1}{l|}{\cellcolor[HTML]{67FD9A}85.71\%} & \cellcolor[HTML]{67FD9A}82.35\% & \multicolumn{1}{l|}{\cellcolor[HTML]{67FD9A}88.10\%} & \cellcolor[HTML]{67FD9A}85.51\% & \multicolumn{1}{l|}{\cellcolor[HTML]{67FD9A}82.14\%} & \cellcolor[HTML]{67FD9A}87.70\% & \multicolumn{1}{l|}{\cellcolor[HTML]{32CB00}97.02\%} & \cellcolor[HTML]{32CB00}96.55\% \\ \hline
\textit{Analytics}     & \multicolumn{1}{l|}{}                                &                                 & \multicolumn{1}{l|}{}                                &                                 & \multicolumn{1}{l|}{}                                &                                 & \multicolumn{1}{l|}{}                                &                                 & \multicolumn{1}{l|}{}                                &                                 & \multicolumn{1}{l|}{}                                &                                 & \multicolumn{1}{l|}{}                                &                                 & \multicolumn{1}{l|}{}                                &                                 \\ \hline
1 Hop                  & \multicolumn{1}{l|}{\cellcolor[HTML]{FFCB2F}61.98\%} & \cellcolor[HTML]{EBFF0C}70.45\% & \multicolumn{1}{l|}{\cellcolor[HTML]{EBFF0C}70.31\%} & \cellcolor[HTML]{EBFF0C}70.16\% & \multicolumn{1}{l|}{\cellcolor[HTML]{FFCB2F}67.71\%} & \cellcolor[HTML]{FFCB2F}65.56\% & \multicolumn{1}{l|}{\cellcolor[HTML]{FFCB2F}68.23\%} & \cellcolor[HTML]{FFCB2F}68.06\% & \multicolumn{1}{l|}{\cellcolor[HTML]{EBFF0C}70.83\%} & \cellcolor[HTML]{EBFF0C}71.13\% & \multicolumn{1}{l|}{\cellcolor[HTML]{FFCB2F}69.71\%} & \cellcolor[HTML]{EBFF0C}70.10\% & \multicolumn{1}{l|}{\cellcolor[HTML]{EBFF0C}76.04\%} & \cellcolor[HTML]{67FD9A}86.39\% & \multicolumn{1}{l|}{-}                               & -                               \\
2 Hop                  & \multicolumn{1}{l|}{\cellcolor[HTML]{EBFF0C}70.83\%} & \cellcolor[HTML]{EBFF0C}75.65\% & \multicolumn{1}{l|}{\cellcolor[HTML]{FFCB2F}67.71\%} & \cellcolor[HTML]{FFCB2F}69.31\% & \multicolumn{1}{l|}{\cellcolor[HTML]{FFCB2F}69.79\%} & \cellcolor[HTML]{EBFF0C}70.10\% & \multicolumn{1}{l|}{\cellcolor[HTML]{FFCB2F}67.19\%} & \cellcolor[HTML]{FFCB2F}68.66\% & \multicolumn{1}{l|}{\cellcolor[HTML]{FFCB2F}67.71\%} & \cellcolor[HTML]{EBFF0C}72.57\% & \multicolumn{1}{l|}{\cellcolor[HTML]{EBFF0C}72.92\%} & \cellcolor[HTML]{EBFF0C}74.51\% & \multicolumn{1}{l|}{\cellcolor[HTML]{EBFF0C}76.04\%} & \cellcolor[HTML]{67FD9A}86.39\% & \multicolumn{1}{l|}{-}                               & -                               \\
3 Hop                  & \multicolumn{1}{l|}{\cellcolor[HTML]{EBFF0C}70.83\%} & \cellcolor[HTML]{EBFF0C}74.07\% & \multicolumn{1}{l|}{\cellcolor[HTML]{EBFF0C}70.31\%} & \cellcolor[HTML]{EBFF0C}71.36\% & \multicolumn{1}{l|}{\cellcolor[HTML]{EBFF0C}71.35\%} & \cellcolor[HTML]{EBFF0C}74.42\% & \multicolumn{1}{l|}{\cellcolor[HTML]{EBFF0C}70.31\%} & \cellcolor[HTML]{EBFF0C}71.07\% & \multicolumn{1}{l|}{\cellcolor[HTML]{FFCB2F}69.79\%} & \cellcolor[HTML]{FFCC67}69.79\% & \multicolumn{1}{l|}{\cellcolor[HTML]{EBFF0C}70.31\%} & \cellcolor[HTML]{EBFF0C}70.16\% & \multicolumn{1}{l|}{\cellcolor[HTML]{EBFF0C}76.04\%} & \cellcolor[HTML]{67FD9A}86.39\% & \multicolumn{1}{l|}{\cellcolor[HTML]{32CB00}96.81\%} & \cellcolor[HTML]{32CB00}96.83\% \\ \hline
\textit{Others}        & \multicolumn{1}{l|}{}                                &                                 & \multicolumn{1}{l|}{}                                &                                 & \multicolumn{1}{l|}{}                                &                                 & \multicolumn{1}{l|}{}                                &                                 & \multicolumn{1}{l|}{}                                &                                 & \multicolumn{1}{l|}{}                                &                                 & \multicolumn{1}{l|}{}                                &                                 & \multicolumn{1}{l|}{}                                &                                 \\ \hline
1 Hop                  & \multicolumn{1}{l|}{\cellcolor[HTML]{EBFF0C}72.50\%} & \cellcolor[HTML]{EBFF0C}73.81\% & \multicolumn{1}{l|}{\cellcolor[HTML]{EBFF0C}71.25\%} & \cellcolor[HTML]{EBFF0C}73.56\% & \multicolumn{1}{l|}{\cellcolor[HTML]{EBFF0C}72.50\%} & \cellcolor[HTML]{EBFF0C}73.17\% & \multicolumn{1}{l|}{\cellcolor[HTML]{EBFF0C}75.00\%} & \cellcolor[HTML]{EBFF0C}76.19\% & \multicolumn{1}{l|}{\cellcolor[HTML]{FFCC67}67.50\%} & \cellcolor[HTML]{FFCC67}68.29\% & \multicolumn{1}{l|}{\cellcolor[HTML]{EBFF0C}71.25\%} & \cellcolor[HTML]{EBFF0C}72.29\% & \multicolumn{1}{l|}{\cellcolor[HTML]{32CB00}90.00\%} & \cellcolor[HTML]{32CB00}94.74\% & \multicolumn{1}{l|}{-}                               & -                               \\
2 Hop                  & \multicolumn{1}{l|}{\cellcolor[HTML]{FFCC67}68.80\%} & \cellcolor[HTML]{FFCC67}69.14\% & \multicolumn{1}{l|}{\cellcolor[HTML]{EBFF0C}76.25\%} & \cellcolor[HTML]{EBFF0C}75.95\% & \multicolumn{1}{l|}{\cellcolor[HTML]{EBFF0C}77.50\%} & \cellcolor[HTML]{EBFF0C}77.50\% & \multicolumn{1}{l|}{\cellcolor[HTML]{EBFF0C}77.50\%} & \cellcolor[HTML]{EBFF0C}76.92\% & \multicolumn{1}{l|}{\cellcolor[HTML]{EBFF0C}76.25\%} & \cellcolor[HTML]{EBFF0C}75.95\% & \multicolumn{1}{l|}{\cellcolor[HTML]{EBFF0C}75.00\%} & \cellcolor[HTML]{EBFF0C}75.00\% & \multicolumn{1}{l|}{\cellcolor[HTML]{32CB00}91.25\%} & \cellcolor[HTML]{32CB00}95.30\% & \multicolumn{1}{l|}{-}                               & -                               \\
3 Hop                  & \multicolumn{1}{l|}{\cellcolor[HTML]{EBFF0C}72.50\%} & \cellcolor[HTML]{EBFF0C}73.17\% & \multicolumn{1}{l|}{\cellcolor[HTML]{EBFF0C}73.75\%} & \cellcolor[HTML]{EBFF0C}74.07\% & \multicolumn{1}{l|}{\cellcolor[HTML]{EBFF0C}75.00\%} & \cellcolor[HTML]{EBFF0C}76.19\% & \multicolumn{1}{l|}{\cellcolor[HTML]{EBFF0C}77.50\%} & \cellcolor[HTML]{EBFF0C}78.05\% & \multicolumn{1}{l|}{\cellcolor[HTML]{EBFF0C}76.25\%} & \cellcolor[HTML]{EBFF0C}77.11\% & \multicolumn{1}{l|}{\cellcolor[HTML]{EBFF0C}77.50\%} & \cellcolor[HTML]{EBFF0C}78.05\% & \multicolumn{1}{l|}{\cellcolor[HTML]{32CB00}90.00\%} & \cellcolor[HTML]{32CB00}94.74\% & \multicolumn{1}{l|}{\cellcolor[HTML]{32CB00}98.75\%} & \cellcolor[HTML]{32CB00}98.59\% \\ \hline \hline
\textbf{Avg.}          & \multicolumn{1}{l|}{\cellcolor[HTML]{EBFF0C}75.38\%} & \cellcolor[HTML]{EBFF0C}77.80\% & \multicolumn{1}{l|}{\cellcolor[HTML]{EBFF0C}79.54\%} & \cellcolor[HTML]{EBFF0C}79.08\% & \multicolumn{1}{l|}{\cellcolor[HTML]{EBFF0C}79.59\%} & \cellcolor[HTML]{EBFF0C}78.82\% & \multicolumn{1}{l|}{\cellcolor[HTML]{67FD9A}80.51\%} & \cellcolor[HTML]{67FD9A}80.06\% & \multicolumn{1}{l|}{\cellcolor[HTML]{EBFF0C}79.21\%} & \cellcolor[HTML]{EBFF0C}78.86\% & \multicolumn{1}{l|}{\cellcolor[HTML]{67FD9A}80.77\%} & \cellcolor[HTML]{67FD9A}80.24\% & \multicolumn{1}{l|}{\cellcolor[HTML]{67FD9A}85.26\%} & \cellcolor[HTML]{32CB00}90.67\% & \multicolumn{1}{l|}{\cellcolor[HTML]{32CB00}97.34\%} & \cellcolor[HTML]{32CB00}97.20\% \\ \hline
\end{tabular}%
}

\label{table:config-experiments-results}
\end{table}
\end{landscape}

\noindent By increasing syntactic structural information (via tokenizing non-terminal nodes), the identifier information present in nodes decreases disproportionately, making it difficult for the model to predict this label. This is also evident by the inverse classification performance between \texttt{Processing} and \texttt{Sharing} labels, i.e., when \textit{Processing} is better predicted (\textit{Exp 1:L\_100}) with tokenized non-terminal nodes, \textit{Sharing} is not and vice versa (\textit{Baseline}). As noted in the baseline, \textit{Sharing} is often implemented in the second and third hops \cite{jain2022pact}. Since we use 100 paths to represent 2 or 3 hops, it can result in an incomplete representation which may not capture relevant paths to identify \texttt{Sharing}. 

Overall, we found an increase in the performance for most cases; thus, for all subsequent experiments, we tokenize non-terminal nodes (and include attention).

\subsubsection{RQ 1.2: LSTM vs. Bi-LSTM Layers}
\label{subsec:lstmvsbilstm}

In comparison to \textit{Exp 1:L\_100}, in \textit{Exp 4:Bi\_100}, we did not observe an increase in the scores when we used Bi-LSTM layers (see Table \ref{table:config-experiments-results}). The average accuracy for \textit{Practice} and \textit{Purpose} labels are 71.34\% and 79.21\% which are similar to those for LSTM layers (71.75\% and 79.54\%, \textit{Exp 1:L\_100}). In both experiments 1 and 4, we used 100 AST paths. These results suggest that an additional backward encoding of AST paths is not very significant, especially when the number of paths is small (i.e., 100). 




However, with a larger number of AST paths, say 200 or 300, Bi-LTSM could potentially improve the scores. We inspect this assumption in RQ 1.3 when we experiment with both LSTM and Bi-LSTM models.

\subsubsection{RQ 1.3: Number of AST Paths}
\label{subsec:num-paths-results}

Overall, we find that with a smaller number of paths (100), LSTM predicts labels with higher accuracy while with a larger number of paths (300), Bi-LSTM performs better. When we use the LSTM layer (i.e., \textit{Exp 1:L\_100}, \textit{Exp 2:L\_200}, and \textit{Exp 3:L\_300}), we notice minimal changes in performance between 100, 200, and 300 AST paths as shown in Table \ref{table:config-experiments-results}. The average accuracy for \textit{Practice} labels are 71.75\%, 71.39\%, and 72.27\%  and for \textit{Purpose} are 79.54\%, 79.59\%, and 80.51\% across these experiments. 
These scores differ within the range of $\sim$1\% which can be attributed to the number of actual paths in a code sample. 

We computed the average number of AST paths in each individual hop and found that the first hop consists of 80 paths, and the second and third hops consist of $\sim$100 paths each. Thus, when we extract, say, 300 paths from a 1\_Hop code sample, more than half of the paths are null paths (i.e., only padding). However, when we extract the same number of paths from a 3\_Hop code sample, where most paths are not null, we observe a slight increase in the scores. For example, in \textit{Exp 3:L\_300}, where we extract 300 paths, \texttt{Collecting\_3\_Hop} has an accuracy of 75.66\% whereas \texttt{Collecting\_1\_Hop} has an accuracy of 72.37\% which matches with our assumption regarding the percentage of null paths in each dataset.

On the other hand, for Bi-LSTM layers, we notice that with a higher number of paths, there is an overall increase in scores. The average accuracy scores for \textit{Practice} are 71.34\%, 72.45\%, and 75.36\% and for \textit{Purpose} are 79.21\%, 80.77\%, and 85.26\%, as shown in Table \ref{table:config-experiments-results} (i.e., \textit{Exp 4:Bi\_100}, \textit{Exp 5:Bi\_200}, and \textit{Exp 6:Bi\_300}). As noted in Section \ref{subsec:lstmvsbilstm}, with only 100 AST paths, the backward encoding of AST paths provides an insignificant increase to the classification scores. However, as we increase the number of paths from 100 to 200 and 300, we observe that this additional encoding noticeably improves the scores. This is also evident based on the increasing differences in accuracy between LSTM and Bi-LSTM layers. For example, the difference between LSTM and Bi-LSTM layers when using 100 AST paths (i.e., \textit{Exp 1:L\_100} vs. \textit{Exp 4:Bi\_100}) for \textit{Practice} labels is less than 1\%. However, when we increase the number of paths to 300, this difference is 3\% with a much larger F-1 score increase of 12\% (i.e., \textit{Exp 3:L\_300} vs. \textit{Exp 6:Bi\_300}). We observe a similar but more conspicuous trend with \textit{Purpose} labels, where the accuracy and F-1 score differences between LSTM and Bi-LSTM layers with 300 paths are $\sim$5\% and $\sim$10\%, respectively. Similar to the LSTM layer's case, using 300 paths for 3\_Hop datasets provides better accuracy. These findings suggest that to represent each code sample, 300 paths is the optimal choice, where each hop contributes to 100 paths. Since each hop on average consists of 100 paths, this number makes sense. 

To summarize the results from RQ 1.1-1.3, we find that the optimal configurations are to: tokenize non-terminal nodes in AST paths when using an attention-based model; use LSTM layers for fewer paths (100) and Bi-LSTM layers when having a greater number of paths (300); and to represent a sample with 100 paths for each method in the sample. We leverage these findings for our multi-head encoder model with attention, where we tokenize each non-terminal node in an AST path. In each encoder head, we use LSTM layers, because each head inputs 100 paths for a single hop.

\subsection{RQ 2: Classification Accuracy}

\begin{figure}[h]
    \centering
    \begin{subfigure}[b]{.26\textwidth}
      \includegraphics[width=\textwidth]{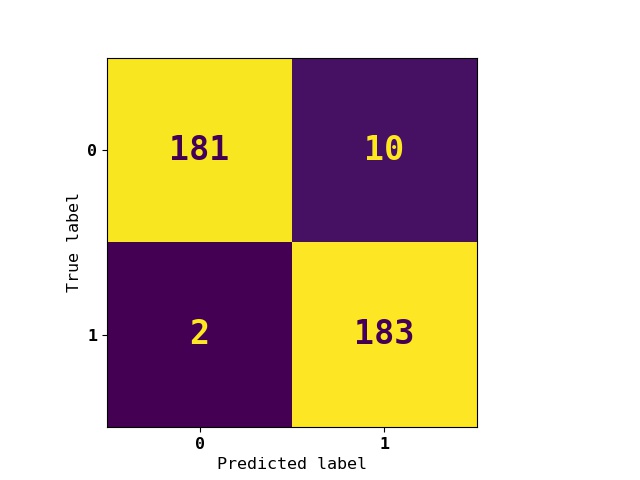}
      \caption{Functionality}
    \end{subfigure}%
    \hskip -6ex
    \begin{subfigure}[b]{.26\textwidth}
      \includegraphics[width=\textwidth]{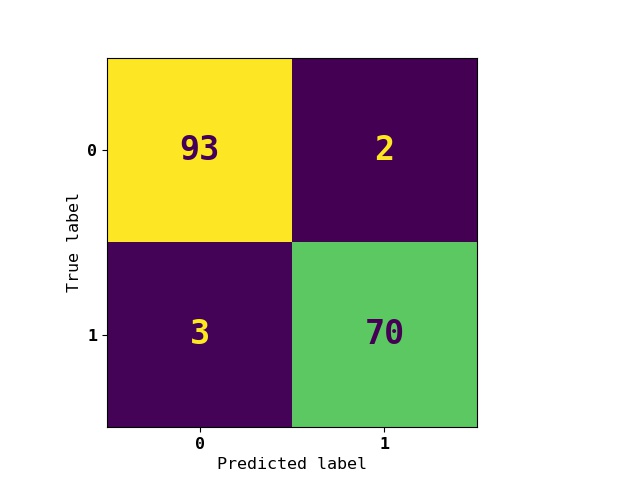}
      \caption{Advertisement}
    \end{subfigure}%
    \hskip -6ex
    \begin{subfigure}[b]{.26\textwidth}
      \includegraphics[width=\textwidth]{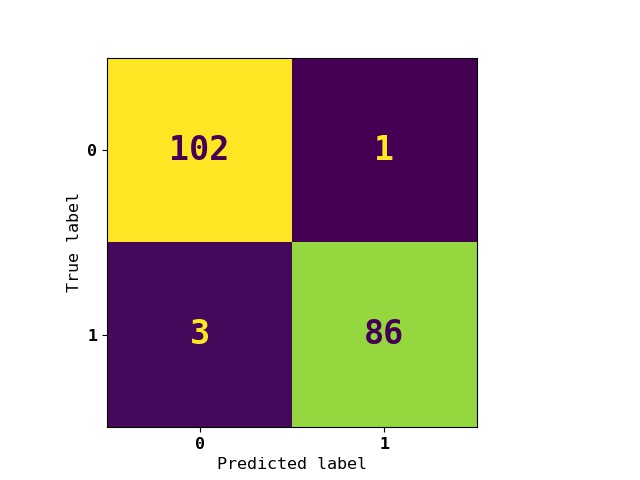}
      \caption{Analytics}
    \end{subfigure}%
    \hskip -6ex
    \begin{subfigure}[b]{.26\textwidth}
      \includegraphics[width=\textwidth]{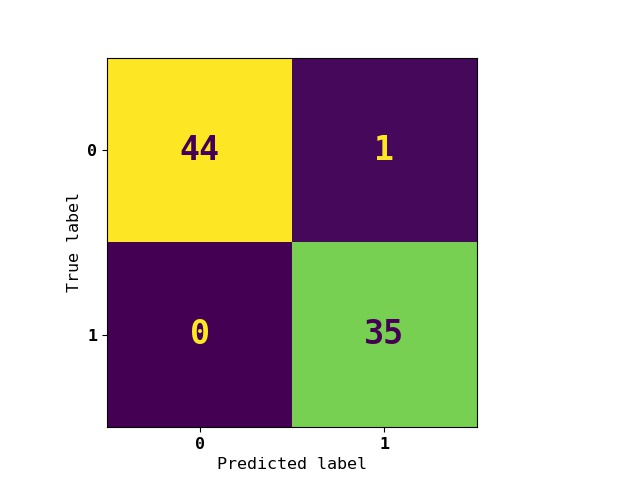}
      \caption{Others}
    \end{subfigure}%
  \caption{Confusion Matrices for \textit{Purpose} labels. $0$ is positive label and $1$ is negative label in each dataset. The x-axis shows the predicted label and the y-axis shows the true label.}
  \label{fig:result-cm-purpose}
\end{figure}

The \textit{Multi-Head Encoder} column of Table \ref{table:config-experiments-results} shows that our model classifies labels with significantly better accuracy than other models. The confusion matrices in Appendix \ref{subsec:appendix-rq-1.2} - Figures \ref{fig:result-cm-practice} and \ref{fig:result-cm-purpose} also demonstrate the unbiased performance of our model for each label. 

We first compare the quantitative scores of our multi-head encoder model with the closest model configuration (\textit{Exp 3:L\_300}). Both models use LSTM layers and an attention module. In Experiment 3, we used 300 AST paths to represent each code sample which is the same number of paths used for the multi-head encoder model (100 AST paths from each hop). The key difference between the two models is their architecture. The average accuracy scores for \textit{Practice} and \textit{Purpose} categories in \textit{Exp 3:L\_300} are 72.27\% and 80.51\%, which are $\sim$22\% and $\sim$17\% lower than that of \textit{Multi-Head Encoder}. This improvement in accuracy is also evident when we compare scores with the \textit{Baseline}, where the increase in average scores for \textit{Practice} and \textit{Purpose} are $\sim$27\% and $\sim$22\%, respectively. 

These significant improvements in the scores can be attributed to the architectural aspects of the model and the optimal configurations. The three encoder heads encode each hop separately, creating better representations and semantically separating them. This is evident from the results for \texttt{Sharing}, for which our model provides a $\sim$20\% increase in accuracy than the closest comparison model (\textit{Exp 3:L\_300}). This is because we use 100 AST paths each to represent second and third hops, which ensures that we capture paths that implement \texttt{`Sharing'}, such as calls to third-party libraries. This was not possible in the baseline approach as noted in RQ 1.1. Furthermore, in each head, the attention module needs to attend over only 100 AST paths which makes it easier for the module to focus on identifiers (which embed the calls to third-party libraries). This is despite the tokenization of non-terminal nodes, which we noted was the probable cause of a decrease in the accuracy for predicting \texttt{Sharing} label in RQ 1.1. Lastly, each head in the multi-head encoder model uses LSTM layers which effectively encode 100 AST paths. Contrarily, in the closest comparison model, the LSTM layers are required to encode 300 paths which is not an optimal configuration. Even if we switch layers in the comparison model to Bi-LSTM, which encodes 300 paths (\textit{Exp 6:Bi\_300}), the results are not nearly as good as the multi-head encoder model.

We also observe that our multi-head encoder model accurately predicts the negative label, as demonstrated by the near diagonal matrices shown in Appendix \ref{subsec:appendix-rq-1.2}, indicating that it predicts without bias towards the positive class.
Similarly, our model provides $\sim$30\% improvement with \texttt{Analytics} label which achieves low classification scores for every other configuration from \textit{Exp 1:L\_100} - \textit{Exp 6:Bi\_300}. Similar to \texttt{Sharing}, \texttt{Analytics} samples also have third-party libraries calling permission-requiring APIs to access personal information, often in the second and third hops. To identify this label, identifier information is necessary. We find that these code samples are often obfuscated, primarily to preserve their methodology for analytics. This obfuscation makes it especially challenging for other models to predict this label. However, with individual attention for each hop, attending over 100 paths, it becomes easier to identify third-party library calls to permission-requiring APIs. 

To summarize, representing each hop as an individual method, preserves semantic differences between three hops and results in significant improvements in the accuracy of the multi-head encoder model. The classification scores of this model are higher than any other model configuration (i.e. \textit{Exp 1:L\_100} - \textit{Exp 6:Bi\_300}).

\subsection{RQ 3: Localization Efficacy}
\label{subsec:localization-results}

Our results show that while there is room for improvement, fine-grained localization is efficacious; that is, it helps in writing privacy statements and can accurately identify statements that implement privacy behaviors. 

\subsubsection{Initial Analysis}
\label{subsubec:init-analysis}

\begin{figure}
    \centering
    \begin{subfigure}[b]{0.5\textwidth}
       \includegraphics[width=\textwidth]{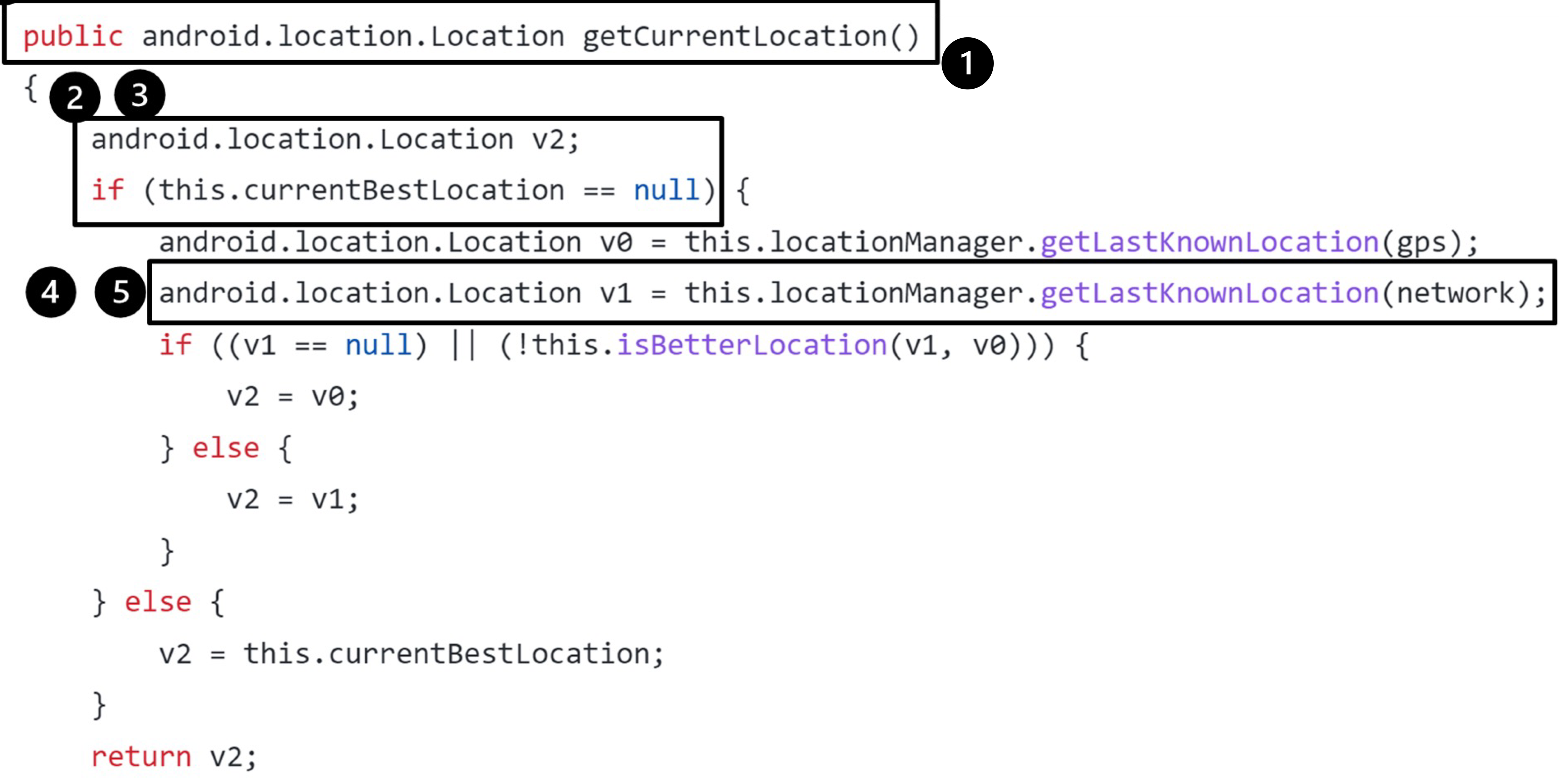}
       \caption{First hop}
   \end{subfigure}
   \begin{subfigure}[b]{0.5\textwidth}
       \includegraphics[width=\textwidth]{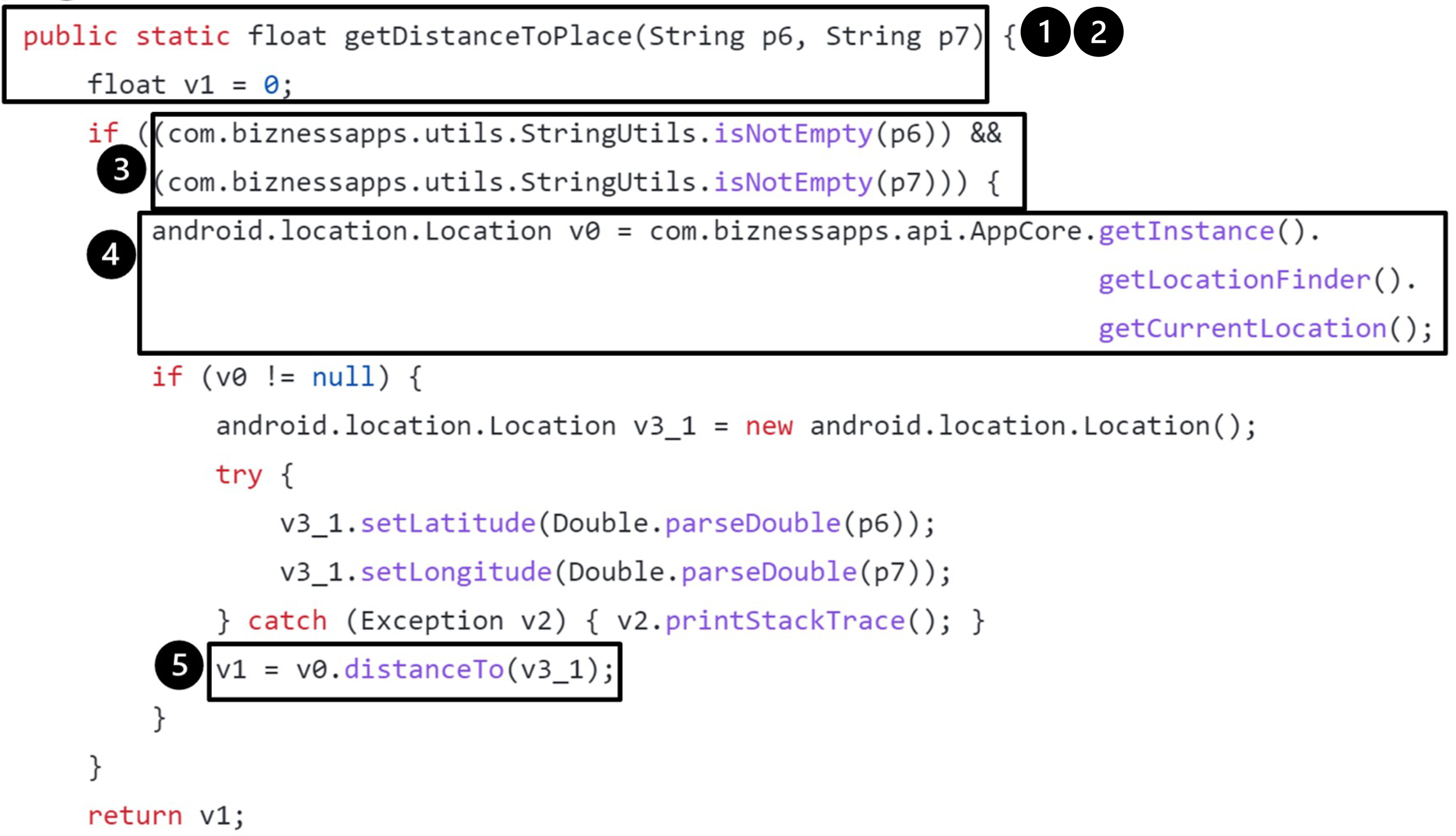}
       \caption{Second hop}
   \end{subfigure}
   \begin{subfigure}[b]{0.5\textwidth}
       \includegraphics[width=\textwidth]{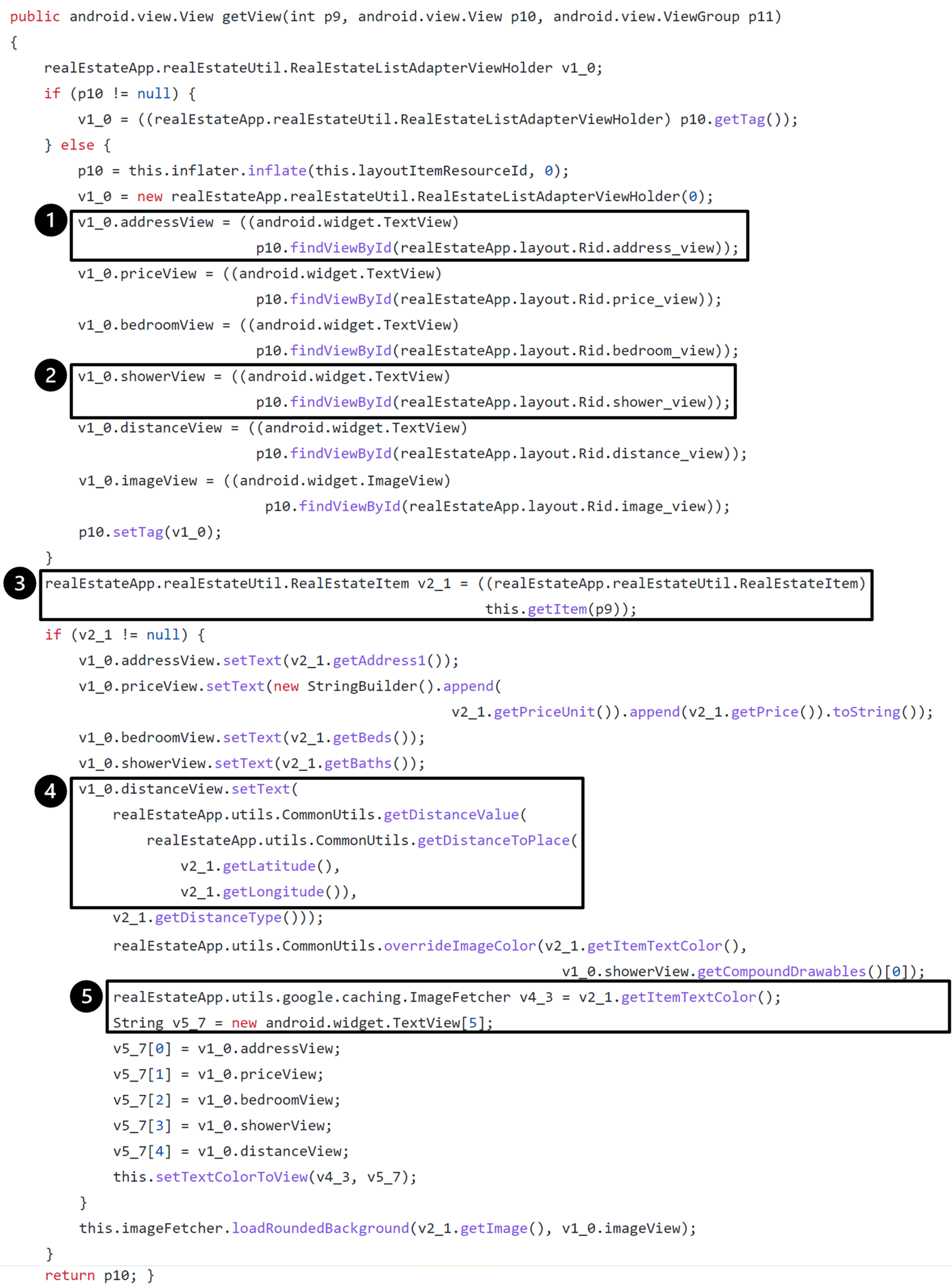}
       \caption{Third hop}
    \end{subfigure}
   \caption{Code snippets and localized statements.}
   \label{fig:result-src-code}
\end{figure}

\begin{figure}
    \centering
    \begin{subfigure}[b]{0.5\textwidth}
       \includegraphics[width=\textwidth]{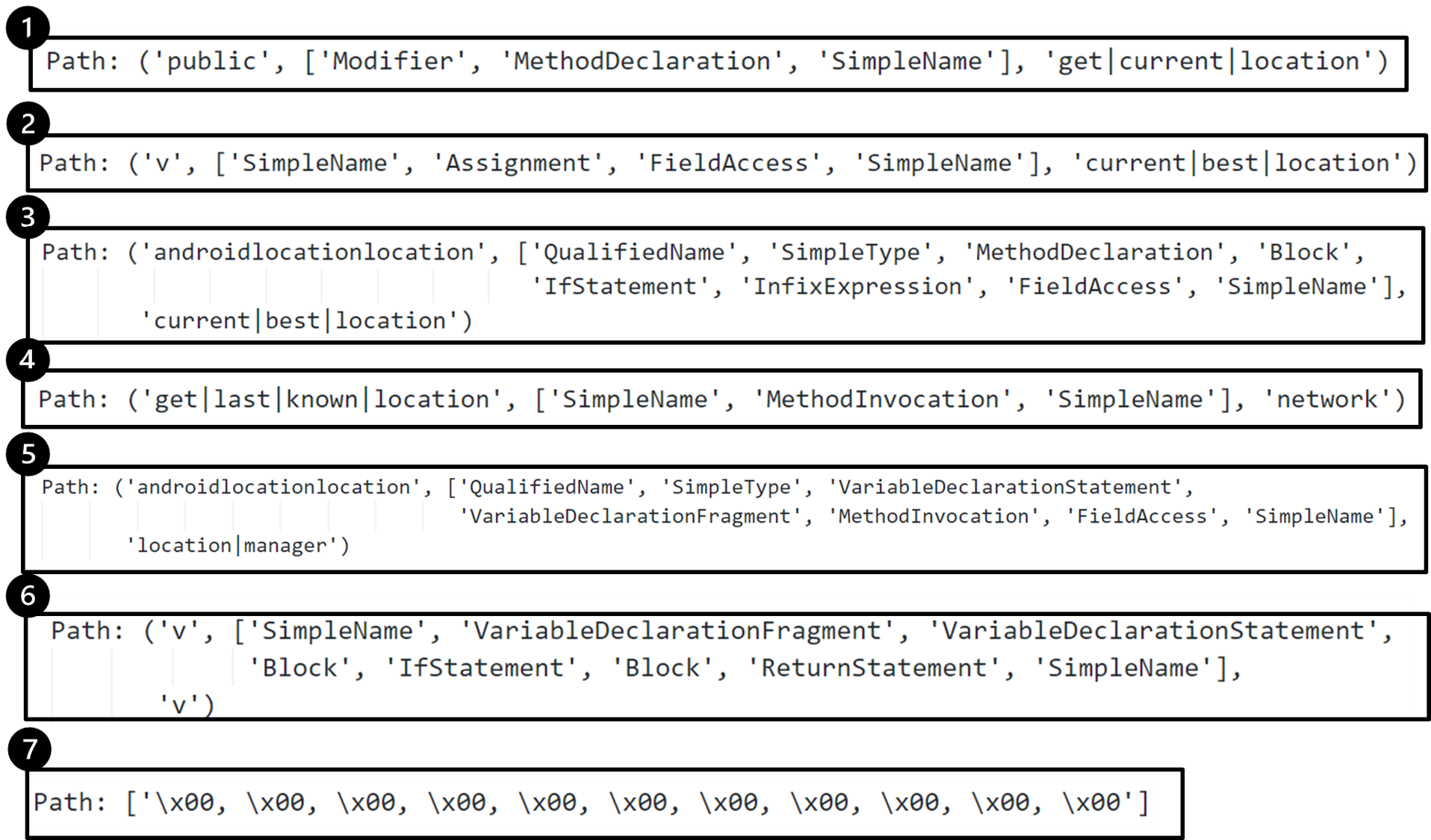}
       \caption{First hop}
   \end{subfigure}
   \begin{subfigure}[b]{0.5\textwidth}
       \includegraphics[width=\textwidth]{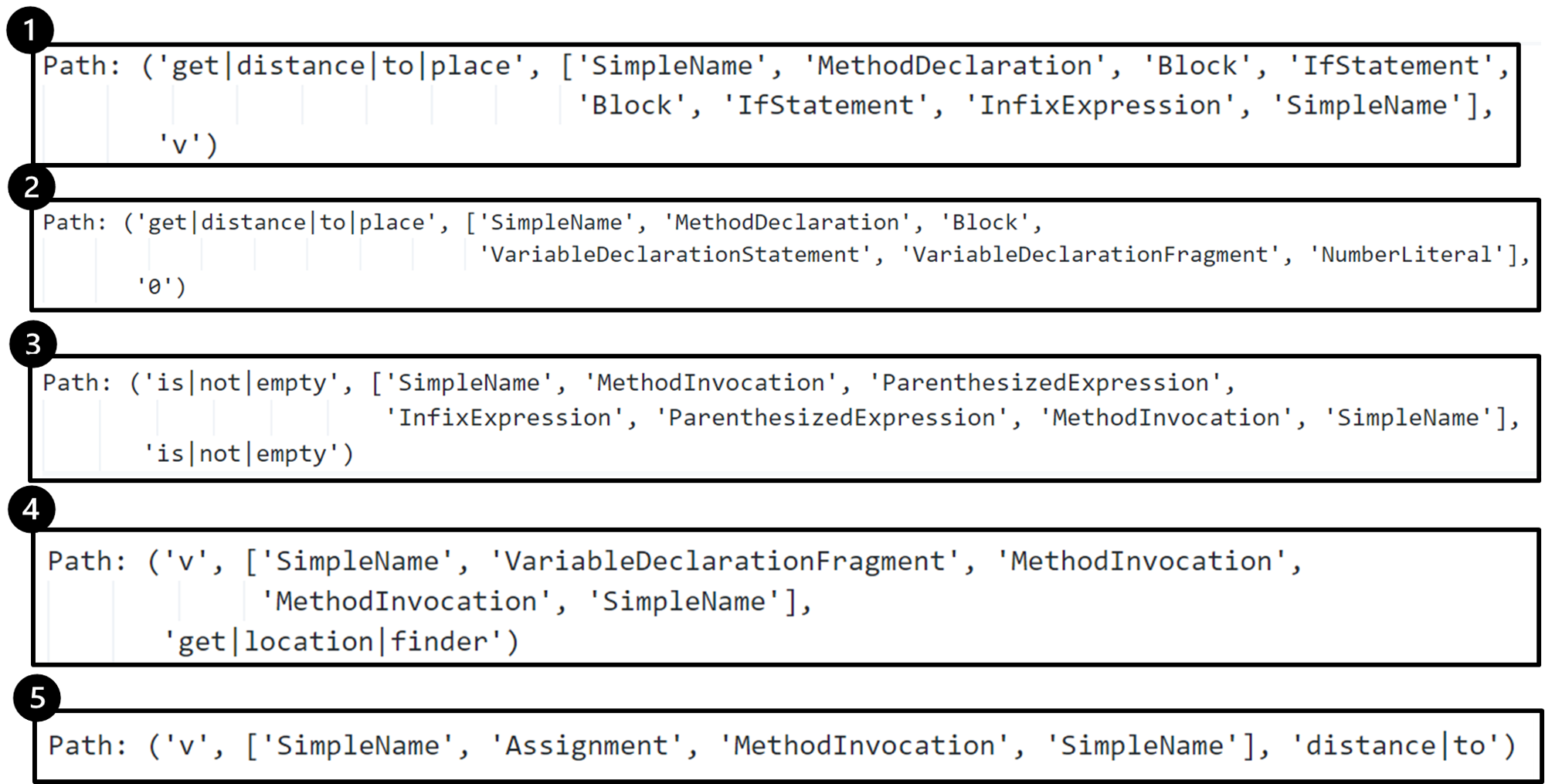}
       \caption{Second hop}
   \end{subfigure}
   \begin{subfigure}[b]{0.5\textwidth}
       \includegraphics[width=\textwidth]{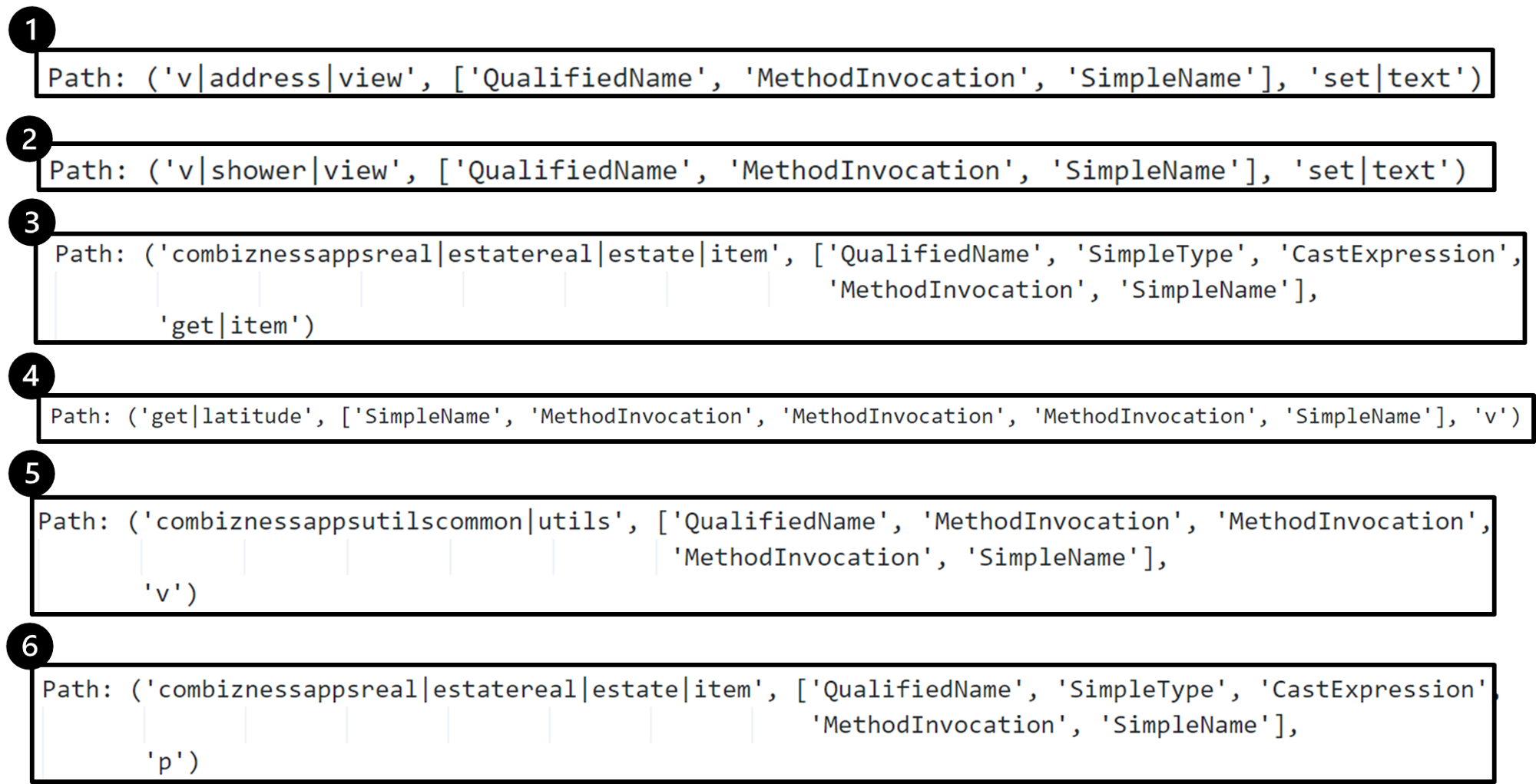}
       \caption{Third hop}
   \end{subfigure}
   \hskip -6ex
   \caption{Most attended AST paths in each hop.}
   \label{fig:result-ast-paths}
\end{figure}

We first examine if our automated approach identifies statements that implement privacy behaviors. Our manual evaluation indicates that our approach succeeds in identifying privacy statements. For instance, consider Figures \ref{fig:result-src-code} and \ref{fig:result-ast-paths} which show an enumerated and highlighted code sample and its AST paths with the highest attention. The code sample gets the location of the user in the first hop (Figure \ref{fig:result-src-code} (a)), calculates its distance to another point (in the second hop (Figure \ref{fig:result-src-code} (b)), and then shows this distance, which is to a real-estate property, in the third hop (Figure \ref{fig:result-src-code} (c)). When we look at the localized statements in the first hop, we observe that most attention is given to statements that get the user's location (i.e., statements 1, 4, and 5 in Figure \ref{fig:result-src-code} (a)), which is the core logic of the first hop and it is implemented in those highlighted statements. These statements are localized based on the terminal nodes of AST paths 1, 4, and 5 shown in Figure \ref{fig:result-ast-paths} (a). Our script does not map AST path 6 shown in the same Figure, since it is obfuscated. Based on the non-terminal nodes `\texttt{IfStatement}' and `\texttt{ReturnStatement}', we can approximate the location to lines 8 or 15. However, it cannot be localized with reasonable certainty; hence, we do not map this path.

For the second hop, focus is given to statements that get the location and calculate its distance to the user's location (i.e., statements 4 and 5 in Figure \ref{fig:result-src-code} (b)). Similar to the first hop, the terminal nodes in AST paths 4 and 5 (Figure \ref{fig:result-ast-paths} (b)) are used to map them. For path 4, the two `\texttt{MethodInvocation}' non-terminal nodes are also used to verify the mapping. This example shows the efficacy of tokenizing non-terminal nodes. Note that in statement 4, the first hop `\texttt{getCurrentLocation}' is highlighted which links the first and second hops together. 

Lastly, in the third hop, most statements focus on developing a view to show details of the property (i.e., statements 1, 2, and 4). In statement 4, the second hop `\texttt{getDistancetoPlace}' is also called, which links the second and third hops together. Method names summarize a method's behavior \cite{alon2018code2seq, jain2022pact} and can be helpful in the comprehension of privacy behaviors. However, this may not always be the case. Consider the method names of the first, second, and third hop in Figure \ref{fig:result-src-code}. The method names in the first two hops are relevant and provide context about how location is being used, i.e., getting the current location and the distance to the current location which are both highlighted in the source code. But, the method name for the third hop does not provide any insight regarding its privacy behavior, which is, thus, not localized in this sample. This suggests that our approach can detect when method names can be helpful and when they cannot (i.e., not localize them). Statements 1, 2, and 4 are spread out, but our approach precisely identifies them without selecting other statements in the same block that are irrelevant. This supports our design decision of statement-level localization. 


Overall, these results indicate that using attention, we can \emph{localize} statements that implement privacy behaviors across three hops as well as the calls to previous hops which is helpful for tracing the flow of information, especially in larger code segments. These results suggest that by mapping highly weighted AST paths, we can provide fine-grained localization of privacy behaviors. 
Figures \ref{fig:result-src-code-2} and \ref{fig:result-ast-paths-2} in Appendix \ref{subsec:appendix-rq-3} show another example of AST paths to source code mappings. 

\subsubsection{RQ 3.1: Analyzing Privacy Statements}

We qualitatively evaluate the dis/advantages of fine-grained localization in helping software professionals write privacy statements. As stated in Section \ref{subsec:experimental_setup}, we ask six software professionals (i.e., annotators) that are divided into two groups to write simple privacy statements for 20 code samples. These two groups are given the same samples, however, the samples that are localized for one group are not localized for the other group and vice-versa. 
The results of this study indicate that fine-grained localization helps annotators easily identify relevant code statements that are necessary for writing privacy statements and saves them the effort and time required to read every line of code to understand the privacy behaviors of the code. The localized statements help annotators with comparatively less experience in software development or privacy to create privacy statements and save up to 74\% of time. 

For each annotator, we discovered negligible differences between the quality of privacy statements for localized and non-localized samples of similar lengths. For example, Annotator \#1 wrote ``We collect device information such as MAC address, Model, OS version, and Serial number.'' for a localized sample and ``We check if your device is connected to a network before loading events.'' for a non-localized one. Both of these statements are written with equal accuracy and detail about the privacy behavior of the code, regardless of fine-grained localization. 
This level of similarity between statements can be attributed to the annotators' knowledge of privacy, which helps them identify privacy behaviors of the code samples even without localization highlights. 

\begin{figure}[htbp]
    \centering
    \fbox{\includegraphics[width=0.39\textwidth, height=4.8cm]{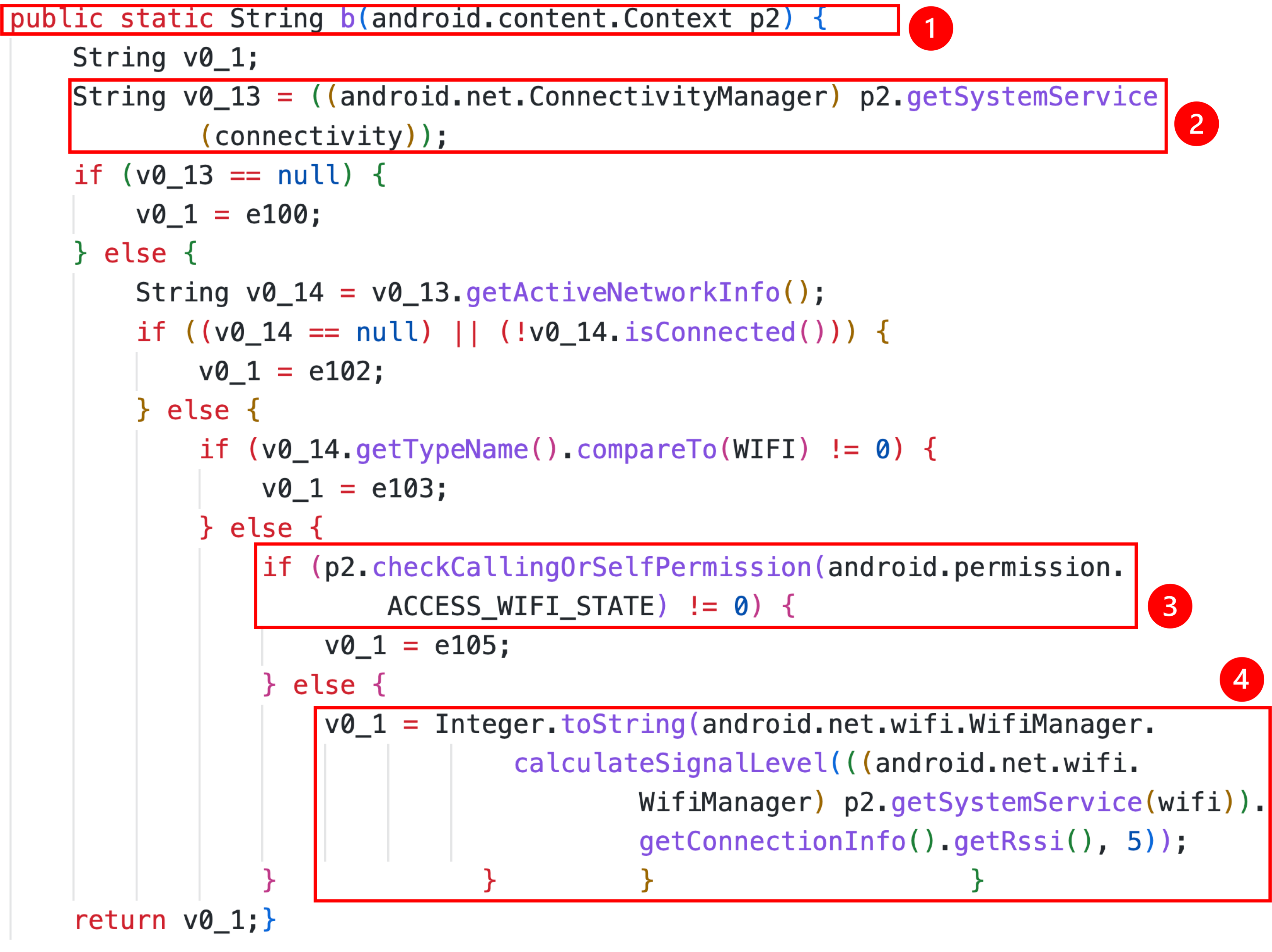}}
    \caption{Appropriate highlights for larger code sample was most helpful for all annotators.}
    \label{fig:result-rq-3.1-long-good}
\end{figure}

We also compared the time taken by annotators to write privacy statements and discovered that while localization highlights saved time for both groups, Group \#2 annotators benefited the most by saving up to 74\% of their time. Annotator \# 4's average time to write a privacy statement, for instance, dropped from 9.7 minutes to 5.5 minutes with localized samples because the highlighted lines helped them by ``drawing [their] attention to privacy-related lines.'' For Group \#1 annotators, we found that localization highlights were more helpful in some scenarios than others. Consider the localization highlights in Figure  \ref{fig:result-rq-3.1-long-good}. Annotator \#2 noted that localization highlights were most helpful since they accurately highlighted ``incidental variable assignments and logic syntax'' in a larger code snippet. These highlights reduced the time and effort required for reading and comprehending privacy behaviors. On the other hand, fine-grained localization was less helpful when there were too many or too few highlights. For example, in Appendix \ref{subsec:appendix-rq-3.1} - Figure \ref{fig:result-rq-3.1-small-bad} almost all statements of a small method are highlighted, rendering localization ineffective. Similarly, a single highlight of a larger method in Appendix \ref{subsec:appendix-rq-3.1} - Figure \ref{fig:result-rq-3.1-long-bad} did not reduce the number of lines required to read and understand privacy behaviors. 

We also compared the time taken by annotators to write privacy statements and discovered that while localization highlights saved time for both groups, Group \#2 annotators benefited the most by saving up to 74\% of their time. Annotator \# 4's average time to write a privacy statement, for instance, dropped from 9.7 minutes to 5.5 minutes with localized samples because the highlighted lines helped them by ``drawing [their] attention to privacy-related lines.'' For Group \#1 annotators, we found that localization highlights were more helpful in some scenarios than others. Consider 

\twocolumn

\onecolumn

\noindent the localization highlights in Figure  \ref{fig:result-rq-3.1-long-good}. Annotator \#2 noted that localization highlights were most helpful since they accurately highlighted ``incidental variable assignments and logic syntax'' in a larger code snippet. These highlights reduced the time and effort required for reading and comprehending privacy behaviors. On the other hand, fine-grained localization was less helpful when there were too many or too few highlights. For example, in Appendix \ref{subsec:appendix-rq-3.1} - Figure \ref{fig:result-rq-3.1-small-bad} almost all statements of a small method are highlighted, rendering localization ineffective. Similarly, a single highlight of a larger method in Appendix \ref{subsec:appendix-rq-3.1} - Figure \ref{fig:result-rq-3.1-long-bad} did not reduce the number of lines required to read and understand privacy behaviors. 

Lastly, we compared the privacy statements between groups for the same samples and discovered Group \#1's statements were either equally good or of better quality compared to those of Group \#2, regardless of localization highlights. For example, Annotator \#3 wrote ``Opens google maps to the devices current location and returns lat/lon when the user clicks'' without localization whereas Annotator \#5 wrote ``We use your location to show your position graphically on Google Maps.'' with localization for the same sample. While the two statements are comparable, the one written by Annotator \#3 has more detail which currently lacks in the statement written by Annotator \#5. These differences in quality and detail can be attributed to the differences in annotators' experiences with privacy and software development. These findings also suggest that developers with limited or no knowledge about privacy could write privacy statements that are comparable to those written by privacy experts when they are provided with localized code samples. However, a more rigorous evaluation will be required to confirm this conclusion, which we plan to do in the future.

\subsubsection{RQ 3.2: Localization Accuracy}
\label{subsubec:localization-acc}

For quantitative evaluation of localization, as mentioned in Section~\ref{subsec:experimental_setup}, we asked the annotators in Group \#1 to label 230 fine-grained localized (highlighted) statements across 20 code samples. These annotators gave a binary label to each localized statement based on whether it implements privacy behavior or not. 

Annotator \#1 found 121 highlights out of 230 ($\sim$52\%) as statements that implement privacy behaviors, while Annotator \#2 found 148 highlights ($\sim$65\%) and Annotator \#3 found 174 statements ($\sim$75\%) as relevant. The inter-annotator agreement scores, i.e., Fleiss's Kappa and Krippendorff's Alpha, among all three annotators were 0.362 for both scores, which for Kappa is considered as ``Fair Agreement". Although our Kappa value is slightly low, other works \cite{harkous2022hark, li2022Understanding, balebako2014privacy} have also achieved low agreement scores since even experts often disagree \cite{balebakoyour2014, li2022Understanding}. We computed the percentages of agreements among annotators for a more thorough analysis. We found that in the best case (where at least one annotator responded `yes'), 85\% of highlighted statements were relevant, and in the average case (where the majority of annotators responded `yes'), 65\% of highlighted statements were relevant. 
This distribution of agreements indicates that all annotators found that the majority of highlighted statements implemented privacy behaviors, where these statements are spread across multiple hops. Their annotations also indicate that our approach has some noise and there are highlighted statements that do not implement privacy behaviors. Thus, there is some scope for improvement in the process of localizing privacy code statements. 

\begin{figure}[htbp]
    \centering
    \fbox{\includegraphics[width=0.475\textwidth]{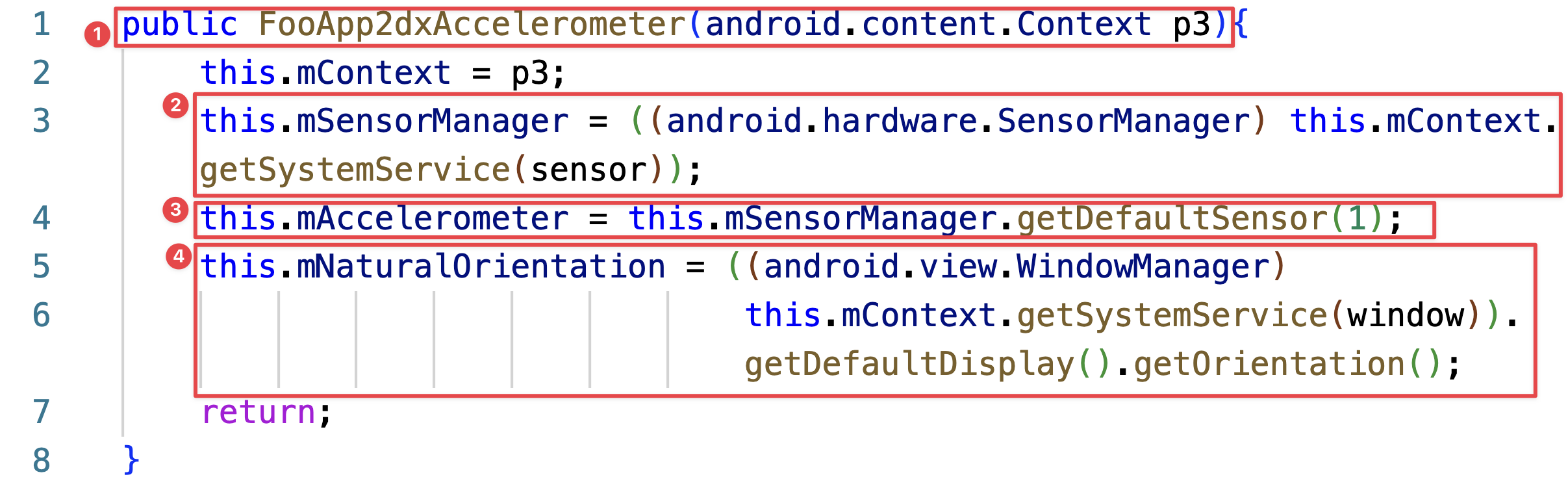}}
    \caption{Disagreement analysis: code snippet 1}
    \label{fig:result-rq3-1}
\end{figure}

We further analyzed the annotation results by inspecting individual cases where annotators disagreed/agreed. We found that, in many cases, annotators agreed when code snippets explicitly used personal information, such as location or email, but disagreed on their subjective view of which statements actually constituted as `implementing privacy behaviors'. Consider the code snippet in Figure \ref{fig:result-rq3-1}, which initializes a \texttt{SensorManager} to access the devices' accelerometer. While Annotators \#2 and \#3 annotated with `yes' for statements 2,3, and 4, Annotator \#1 disagreed. For statement 1, Annotator \#3 responded with a `yes' but Annotators \#1 and \#2 disagreed since these statements provide \textit{implicit} access to a personal accelerometer (which may not be considered as personal information by some). These demonstrate the subjectivity of privacy that exists even between experts which is reported in literature \cite{balebakoyour2014, li2022Understanding}. Although the code snippet in Figure \ref{fig:result-rq3-1} does not explicitly consume the sensor information, the method name and the parameters used can help understand \textit{how} and \textit{why} the information is used. 

\begin{figure}[htbp]
    \centering
    \fbox{\includegraphics[width=0.475\textwidth]{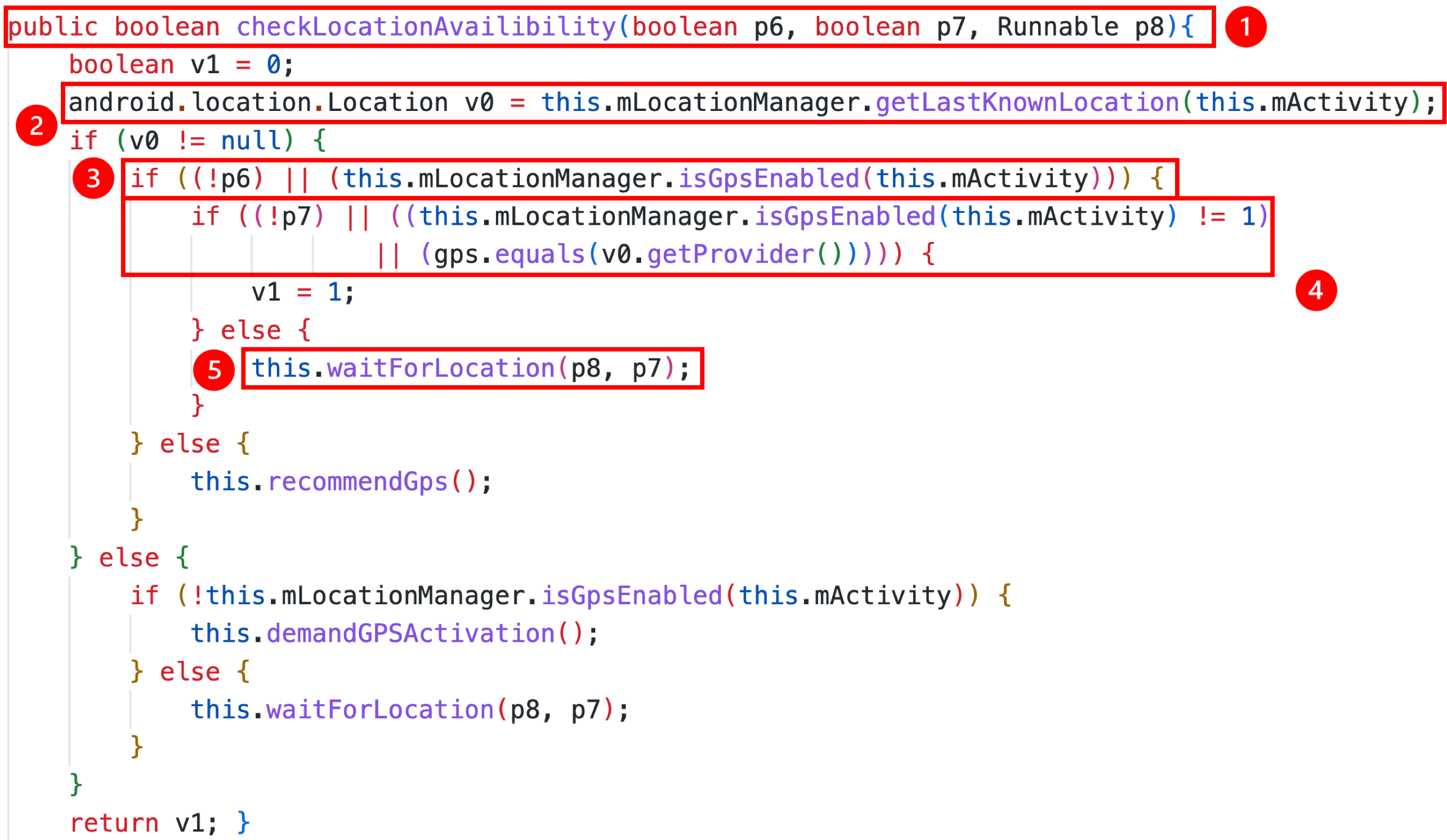}}
    \caption{Statements explicitly using personal information (e.g., 2-5), implement privacy behaviors.}
    \label{fig:result-rq3.1-1}
\end{figure}

Consider another example in Appendix \ref{subsec:appendix-rq-3.2} - Figure \ref{fig:result-rq3-2}, which shows a code snippet from an advertisement library using personal information. Here, all three annotators found the use of `traditional` personal information as implementing privacy behaviors and completely agreed on the highlights. Specifically, they agreed that highlights 4-6 and 8-11 localize privacy behaviors. This example demonstrates that our approach identifies the use of sensitive personal information that is not captured by API calls. For example, date of birth, gender, income, and ethnicity are highly sensitive personal information that are not provided via API calls. It is to be noted that the code samples in the dataset are \textit{not} annotated with any localization information and therefore, all predictions about privacy-relevant statements are learned by the model on its own. Hence, our approach is able to identify such privacy behaviors and can help developers provide much granular insight into their implementations.

\label{sec:results}

\section{Limitations}

We plan to address the following limitations in future.

\noindent \textbf{Automated mapping:} One challenge of our automated approach is to provide an exact match of every path with a statement in the source code. Since a path is the traversal of a sub-tree, it may contain terminal and non-terminal nodes of different statements, making it difficult to provide a one-to-one mapping. In some cases, code samples and thus, the terminal nodes are obfuscated which makes mapping challenging. Thus, our script sometimes maps one path to several statements (i.e., one-to-many), since it only matches terminal nodes. To mitigate this challenge, we manually verify and correct the automated mappings based on non-terminal nodes (such as using \texttt{IfExpression} tokens). If we cannot map the paths with reasonable certainty, we do not map them at all. In the future, we will revise our model architecture to provide source code tokens along with the Abstract Syntax Tree (not paths of AST) which will allow the model to map terminal and non-terminal nodes to source code tokens thereby eliminating the need for separate scripts for mapping. While our approach does not always map obfuscated nodes, developers who will use our work will have access to unobfuscated code. 

\noindent \textbf{Detecting privacy statements:} Our approach accurately identifies privacy-relevant statements in source code in a majority of cases; however, there are instances, when less relevant paths were attended. For example, in the second hop of Figure \ref{fig:result-src-code} (b), the model focuses on statements that check whether input parameters are empty (i.e., statements 3 in the figure). The model also attends null paths (i.e., padding), such as statement 7 in Figure \ref{fig:result-ast-paths} (a)). As with any machine learning approach, there are false positive predictions. In the future, we plan to minimize false positive localization by using our revised model which will utilize the contextual information from source code tokens and syntactic information from the AST for localization. We will also use attention masks to allow model to differentiate between padding and true AST paths. 

\noindent \textbf{Evaluation:} We evaluated our work with 20 samples which may seem trivial, but the ADPAc dataset is prepared from ~15,000 real Android applications \cite{jain2022pact}. Moreover, we extracted the 20 samples from the test sets of ADPAc dataset which guaranteed that they were not seen during the training phase, thereby evaluating our work on real-world application code. Furthermore, in our qualitative evaluation, we selected annotators with some experience in software development and privacy. However, they are not the authors of the source code they evaluated, nor had access to the entire app's source code which may have impacted accuracy for some samples. 

\label{sec:limitations}

\section{Conclusions and Future Work}

In this paper, we described a novel approach to provide fine-grained localization of privacy behaviors in an application's source code for generating privacy labels and helping developers write privacy statements. We developed a novel multi-head encoder model that creates individual representations of multiple methods and then uses attention to identify relevant statements in those representations. 
To identify optimal model configurations, we first conducted six sets of experiments and then trained our model using the optimal configurations. Next, we used our model to predict \textit{Privacy Action} labels. 
Our quantitative results indicate that our unique architecture significantly outperforms the baseline, and achieves high classification accuracy scores of 91.41\% - 98.45\% in predicting \textit{Privacy Action} labels. We also evaluated our fine-grained localization approach manually as well as with six software professionals. Manual evaluation results indicate that our automated approach correctly highlights privacy-relevant statements in most cases, but may need manual curation to remove the false positives mappings. We plan to address this challenge in the future and we discuss this in in Section~\ref{sec:limitations}.

Our qualitative evaluation demonstrates that our approach helps professionals easily identify relevant code statements that are necessary for writing privacy statements and saves them the effort and time required to read every line of code to understand their privacy behaviors. The time required is reduced up to 74\% for professionals with lower expertise. 
We also evaluated the accuracy of our approach with three of the six  professionals with more expertise and found that our model identifies relevant statements that implement privacy behaviors in the majority of the cases. Based on these results, our approach provides a mechanism for fine-grained localization of privacy behaviors. In this work, we only demonstrated the feasibility of fine-grained localization with six software professionals at our university. 
In the future, we will conduct a user study with several Android application developers to evaluate our localization approach's efficacy, and also perform a comparative analysis on the quality and timing of the privacy statements creation.

\label{sec:conclusions}

\section*{Data Availability}

Our datasets, models, and training scripts are available on the GitHub page of our project at \url{https://github.com/PERC-Lab/Fine_Grained_Localization}

\label{sec:data-avail}

\section*{Acknowledgements}

We would like to thank Sam Morse, Mac Creamer, Max Prybylo, Adam Green, Sean Radel, Zack Delile, and Theo Brucker for their contributions to this work.

\bibliographystyle{unsrt}  
\bibliography{main}

\newpage

\section{Abstract Syntax Tree and AST Paths}
\label{subsec:appendix-ast-paths}

\begin{figure}[ht]
    \centering
    \subfloat[][Code Sample]{{\includegraphics[width=0.48\textwidth]{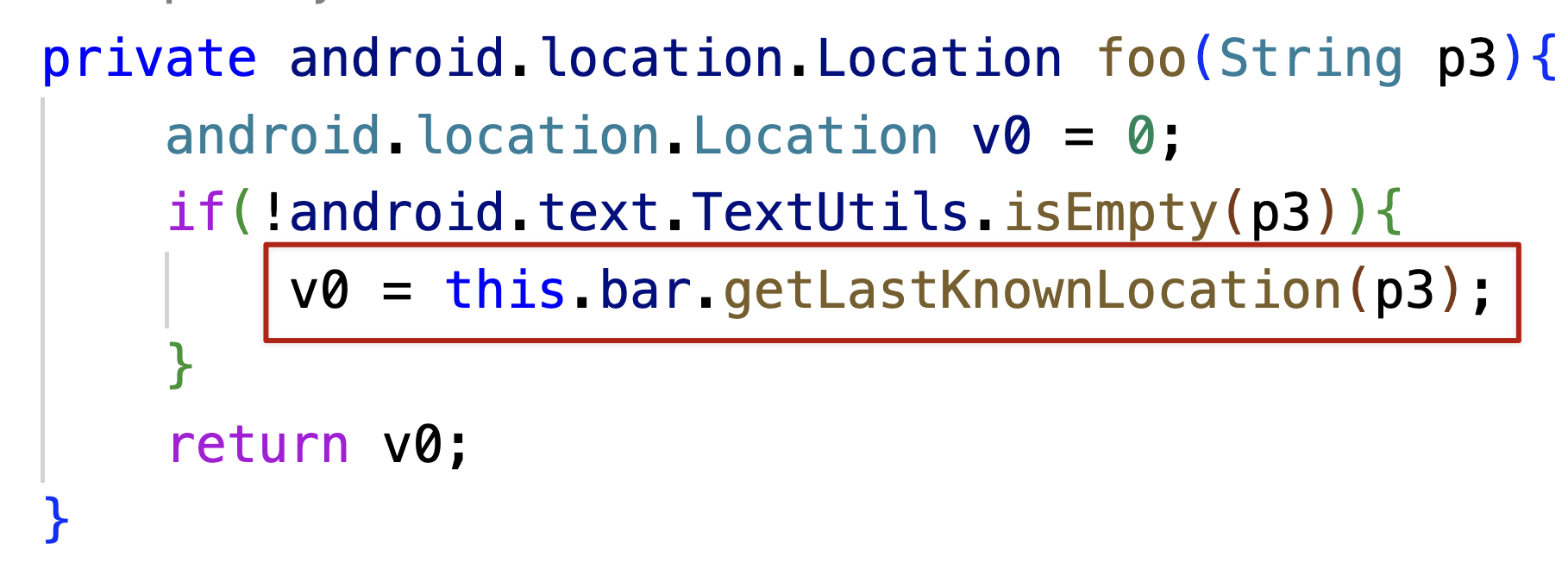}}}\hfill
    \subfloat[][Abstract Syntax Tree (AST)]{{\includegraphics[width=0.4\textwidth]{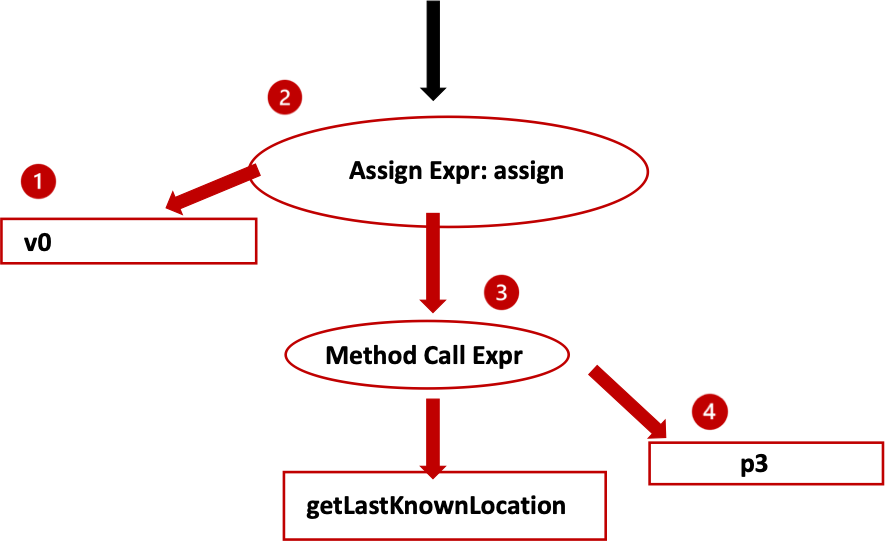}}}\hfill
    \subfloat[][AST Paths]{{\includegraphics[width=0.4\textwidth]{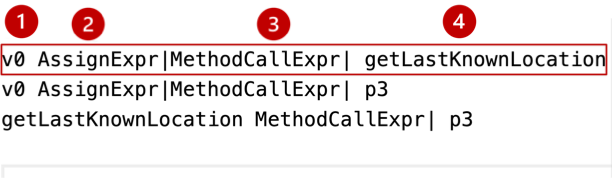}}}\hfill
    \caption{(a) A Code Snippet, (b) Its Partial AST, and (c) Corresponding AST Paths. The Partial AST in (b) Represents the Syntactic Information of the Code Segment Highlighted in Green in (a)}
    \label{fig:ast}
\end{figure}

Source code can be represented as abstract syntax trees (AST) to show their syntactic structure.  Figures \ref{fig:ast} (a) and (b) show a code segment and its partial AST. The leaves of an AST, which are identifiers such as variables and names, are called \textit{terminal nodes} (rectangular nodes in Figure \ref{fig:ast} (b)). The non-leaves, which represent the syntactic structures such as if-statements and loops, are called \textit{non-terminal nodes} \cite{alon2018code2seq} (oval nodes in Figure \ref{fig:ast} (b)). Traversing from one terminal node to another is referred to as an AST path. Figure \ref{fig:ast} (c) shows a list of AST paths traversed from the partial AST in Figure \ref{fig:ast} (b). Since an AST contains useful syntactic information about a code snippet, recent work in code summarization \cite{haque2020improved, alon2018code2seq, hu2018deep} use AST paths to represent code. ADPAc contains the AST paths of code samples and their labels which we use in this work. 

\newpage




\section{RQ 1.2: Confusion Matrices}
\label{subsec:appendix-rq-1.2}


\begin{figure}[h]
    \centering
    \begin{subfigure}[b]{.26\textwidth}
      \includegraphics[width=\textwidth]{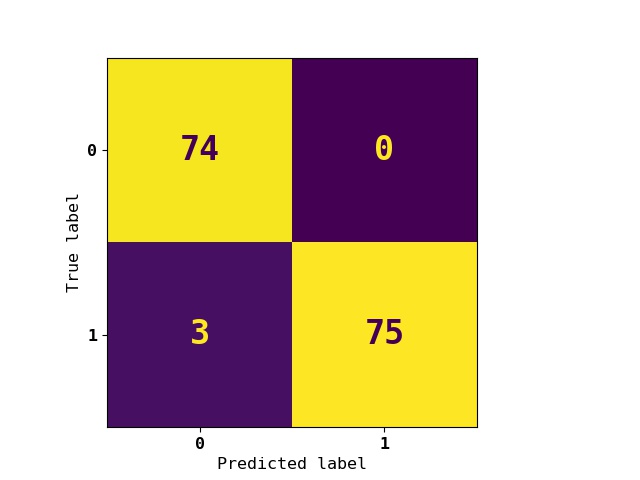}
      \caption{\texttt{Collecting}}
    \end{subfigure}%
    \hskip -6ex
    \begin{subfigure}[b]{.26\textwidth}
      \includegraphics[width=\textwidth]{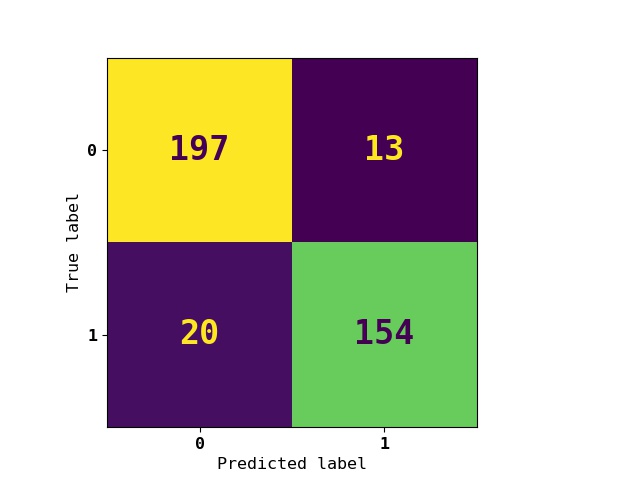}
      \caption{\texttt{Sharing}}
    \end{subfigure}%
    \hskip -6ex
    \begin{subfigure}[b]{.26\textwidth}
      \includegraphics[width=\textwidth]{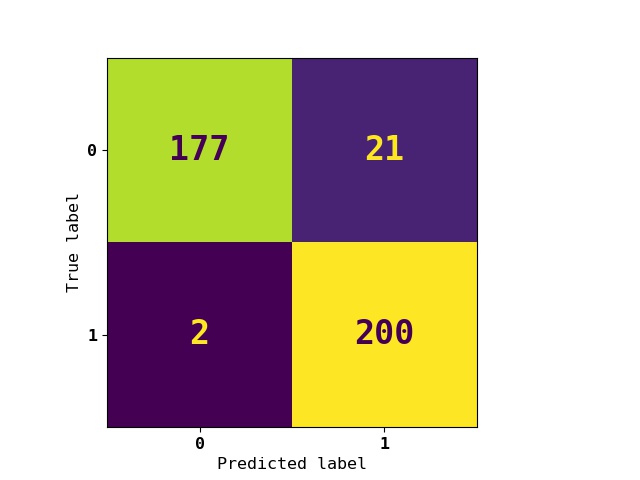}
      \caption{\texttt{Processing}}
    \end{subfigure}%
    \hskip -6ex
    \begin{subfigure}[b]{.26\textwidth}
      \includegraphics[width=\textwidth]{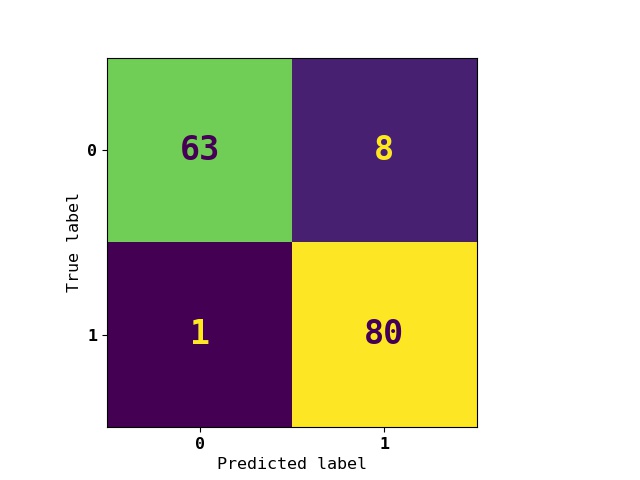}
      \caption{\texttt{Others}}
    \end{subfigure}%
  \caption{Confusion Matrices for \textit{Purpose} labels. $0$ is positive label and $1$ is negative label in each dataset. The x-axis shows the predicted label and the y-axis shows the true label.}
  \label{fig:result-cm-practice}
\end{figure}

\section{RQ 1.3: Attention Maps}
\label{subsec:appendix-rq-1.3}


\begin{figure}[!h]
    \centering
    \begin{subfigure}[b]{0.27\textwidth}
      \includegraphics[width=\textwidth]{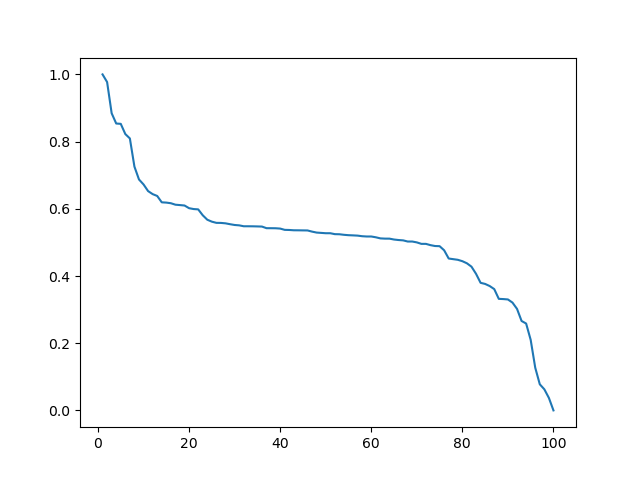}
      \caption{First Hop}
  \end{subfigure}
  \hskip -4ex
  \begin{subfigure}[b]{0.27\textwidth}
      \includegraphics[width=\textwidth]{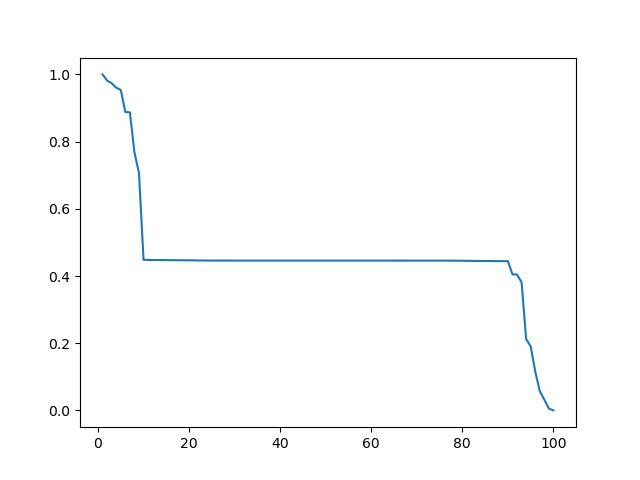}
      \caption{Second Hop}
  \end{subfigure}
  \hskip -4ex
  \begin{subfigure}[b]{0.27\textwidth}
      \includegraphics[width=\textwidth]{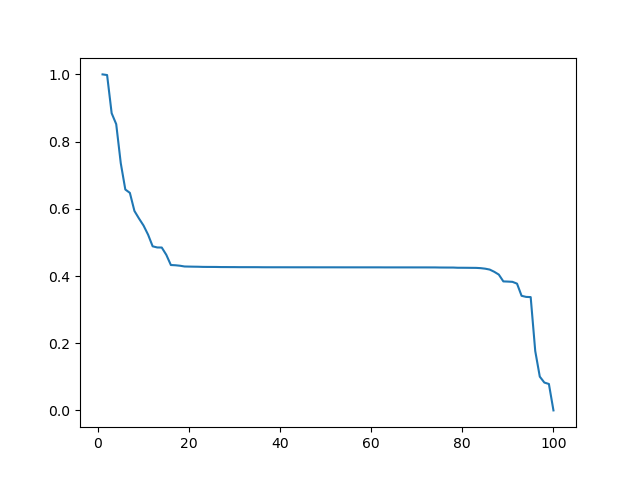}
      \caption{Third Hop}
    \end{subfigure}
  \caption{Attention Maps of Individual Hops for Selected Code Sample.}
  \label{fig:result-attentn-map}
\end{figure}

\newpage

\twocolumn

\section{RQ 3: Localization Feasibility -- Additional Examples}
\label{subsec:appendix-rq-3}

\hskip -6ex

\begin{figure}[t]
    \begin{subfigure}[b]{0.5\textwidth}
      \includegraphics[width=\textwidth]{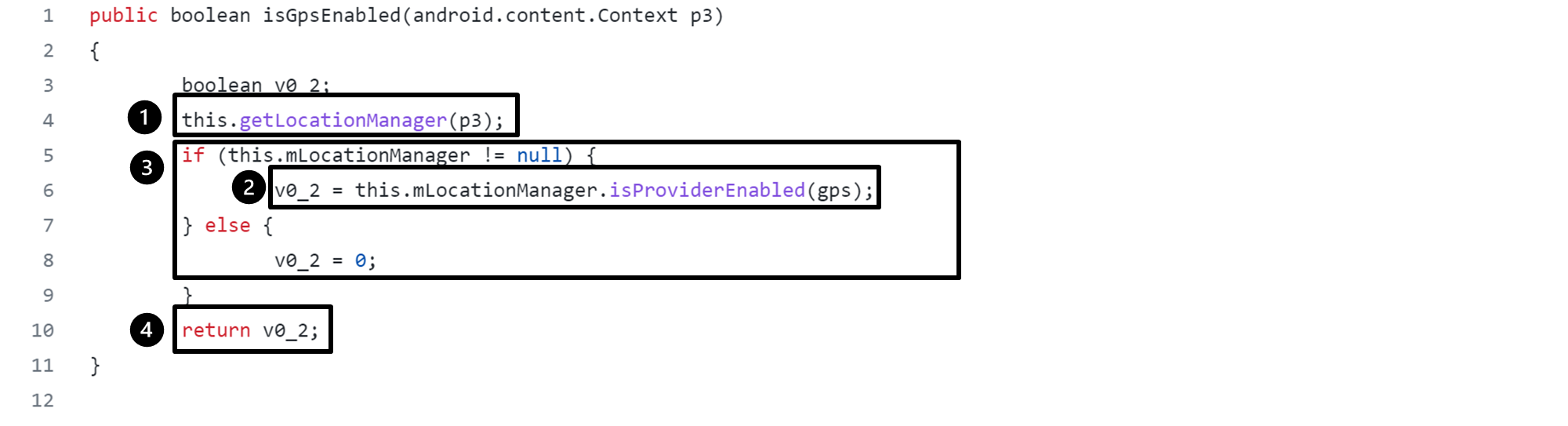}
      \caption{First Hop}
  \end{subfigure}

  
  \begin{subfigure}[b]{0.5\textwidth}
      \includegraphics[width=\columnwidth]{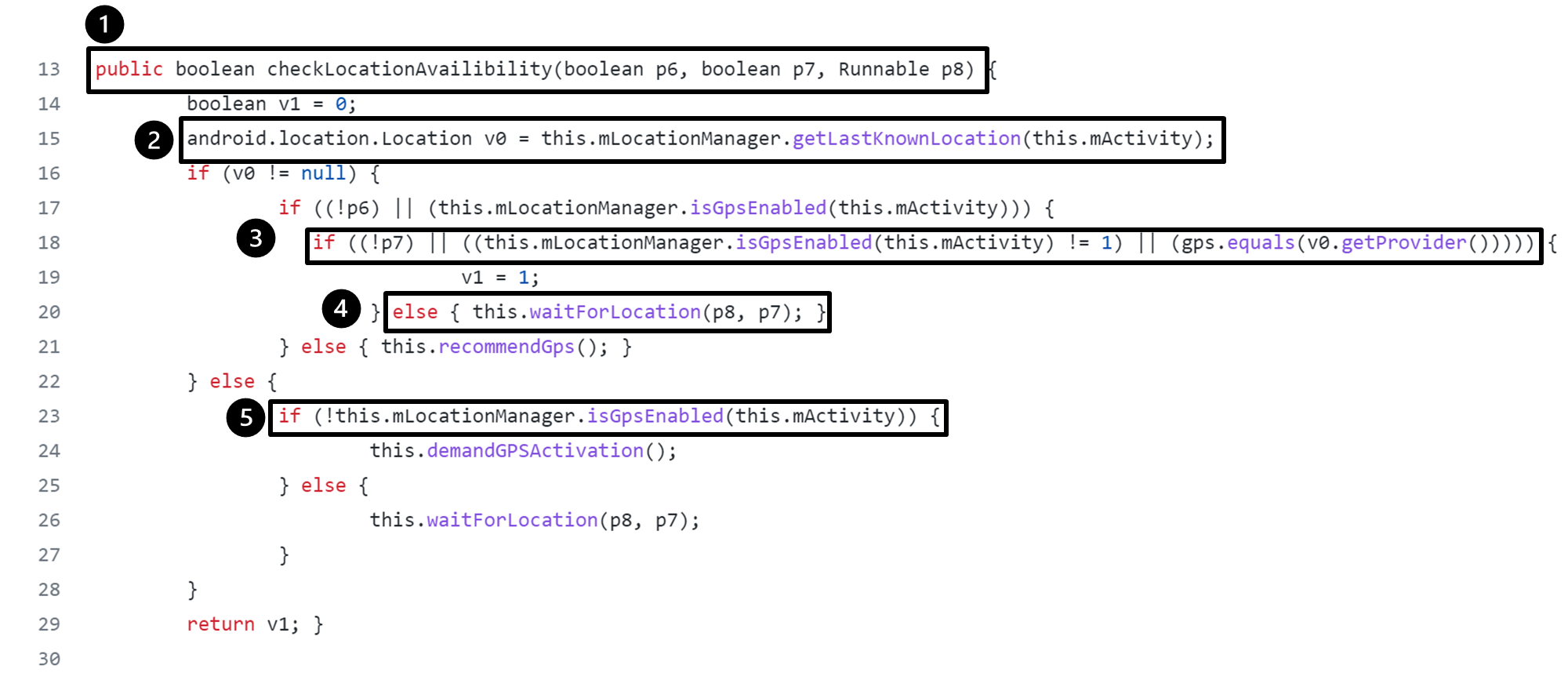}
      \caption{Second Hop}
  \end{subfigure}
  \hskip -6ex \\
  \begin{subfigure}[b]{0.5\textwidth}
      \includegraphics[width=\textwidth]{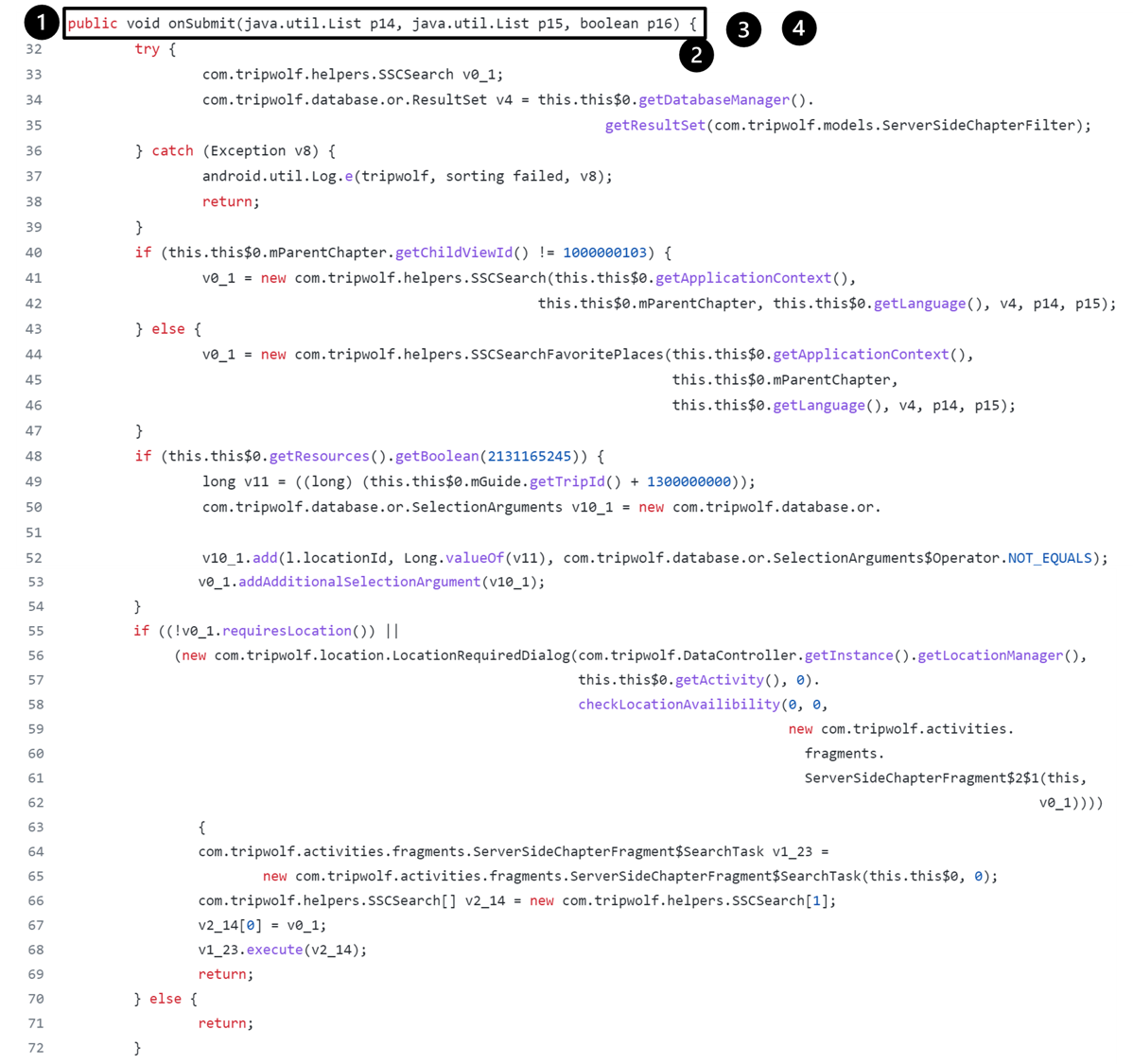}
      \caption{Third Hop}
    \end{subfigure}
  \caption{Code Snippets and Localized Statements of Selected Code Sample.}
  \label{fig:result-src-code-2}
\end{figure}

\begin{figure}[h!]
    \begin{subfigure}[b]{0.5\textwidth}
      \includegraphics[width=\textwidth]{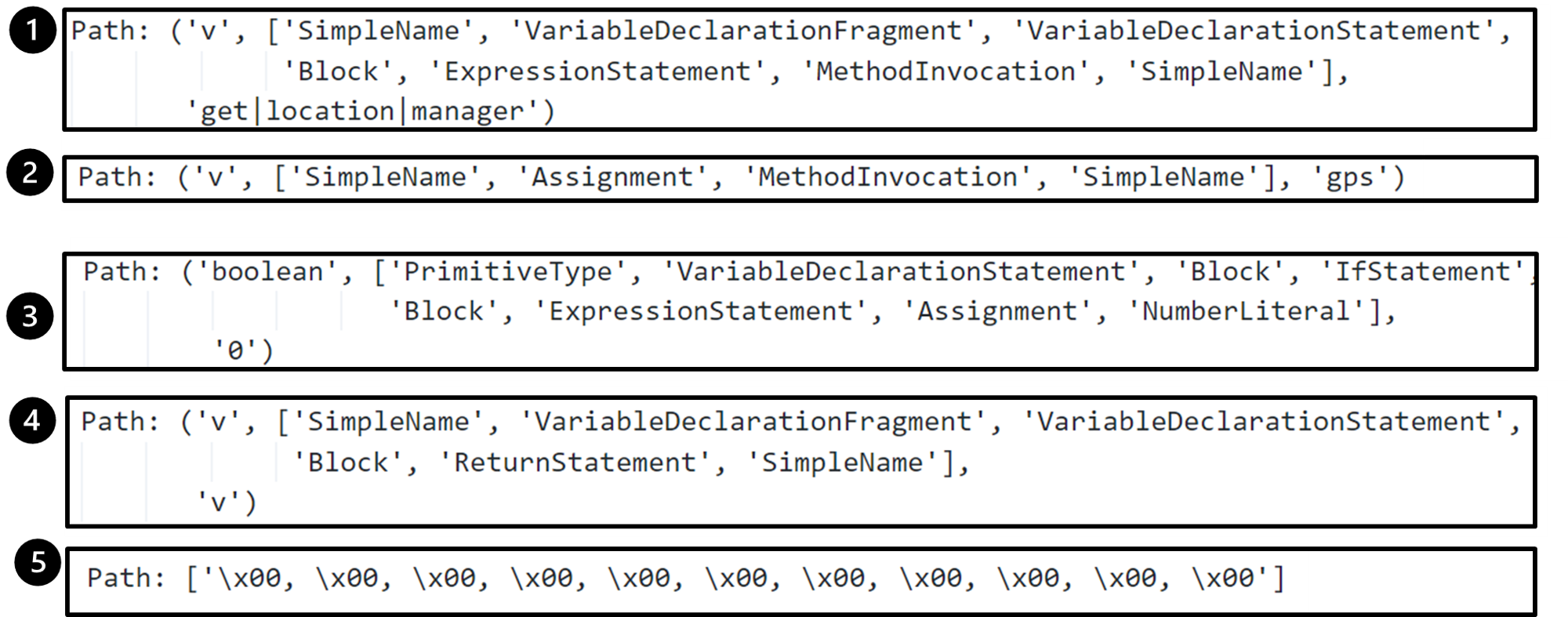}
      \caption{First Hop}
  \end{subfigure}
  \begin{subfigure}[b]{0.5\textwidth}
      \includegraphics[width=\textwidth]{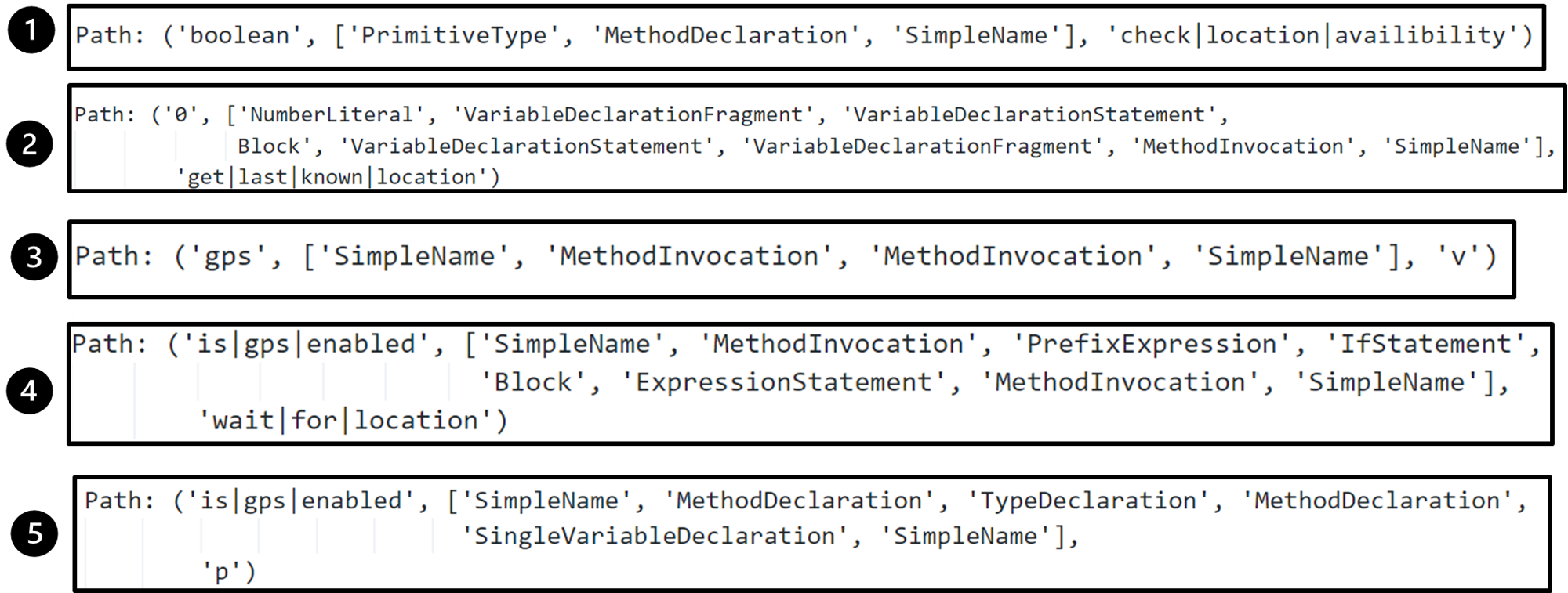}
      \caption{Second Hop}
  \end{subfigure}
  \begin{subfigure}[b]{0.5\textwidth}
      \includegraphics[width=\textwidth]{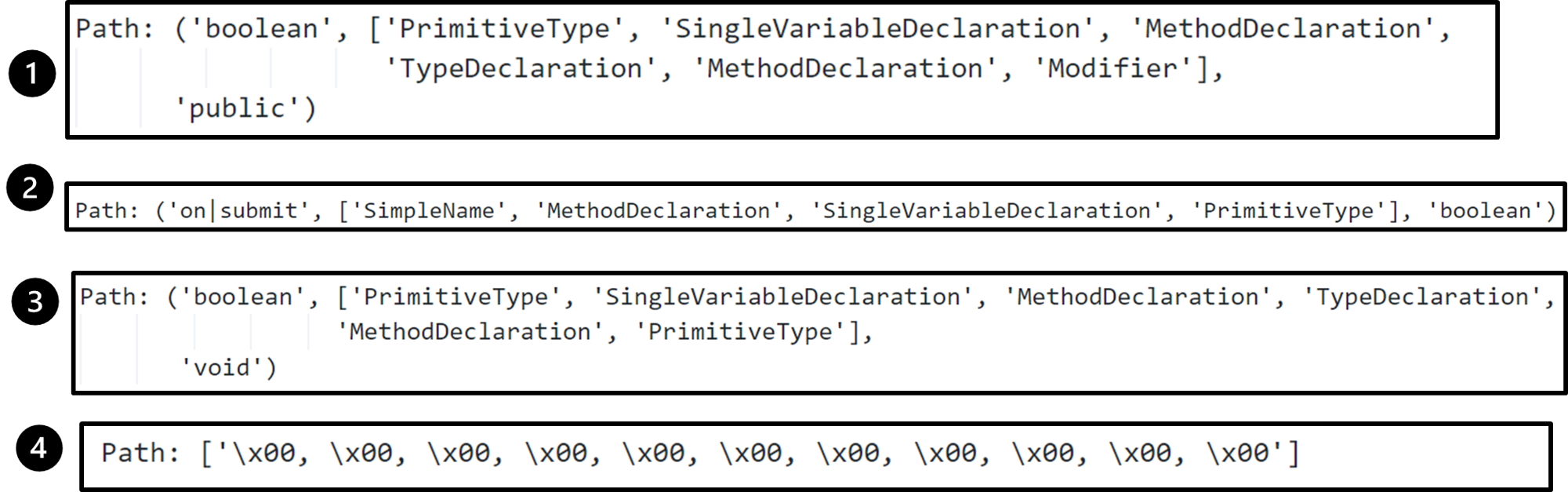}
      \caption{Third Hop}
  \end{subfigure}
  \caption{Most Attended AST Paths in Each Hop for Selected Code Sample.}
  \label{fig:result-ast-paths-2}
\end{figure}

\onecolumn

\clearpage

\section{RQ 3.1: Analyzing Privacy Statements -- Examples}
\label{subsec:appendix-rq-3.1}

\begin{figure}[htbp]
    \centering
    \fbox{\includegraphics[width=0.475\textwidth]{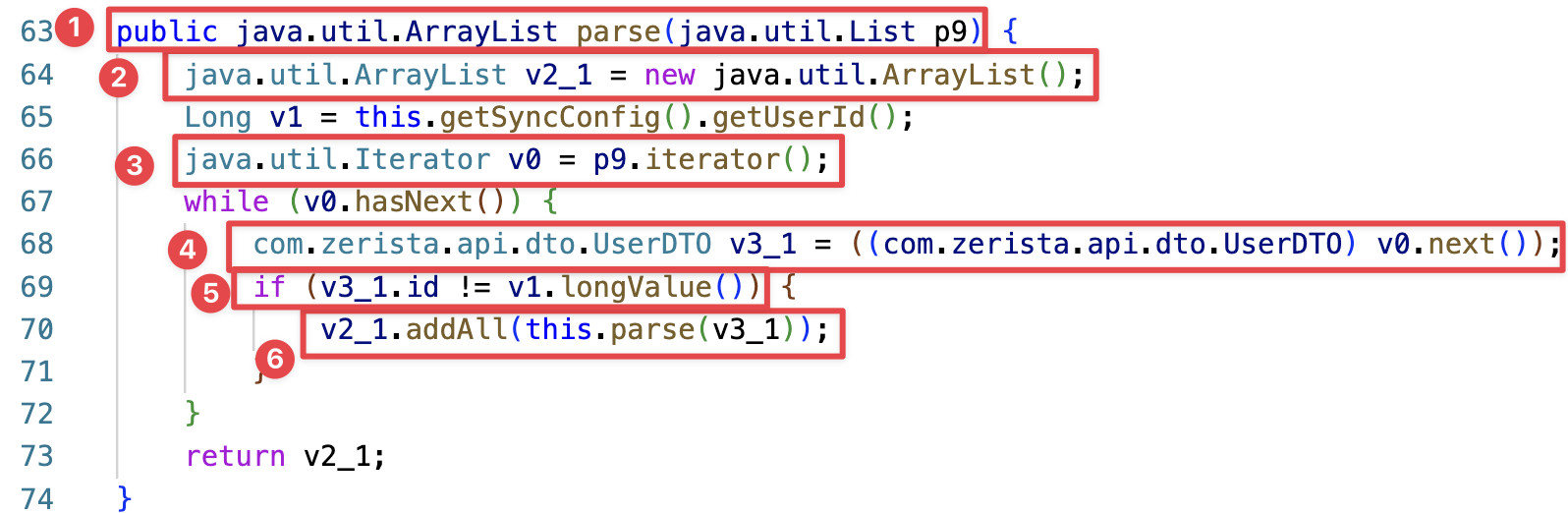}}
    \caption{Too many highlights for small code samples rendered localization ineffective.}
    \label{fig:result-rq-3.1-small-bad}
\end{figure}

\begin{figure}[htbp]
    \centering
    \fbox{\includegraphics[width=0.475\textwidth]{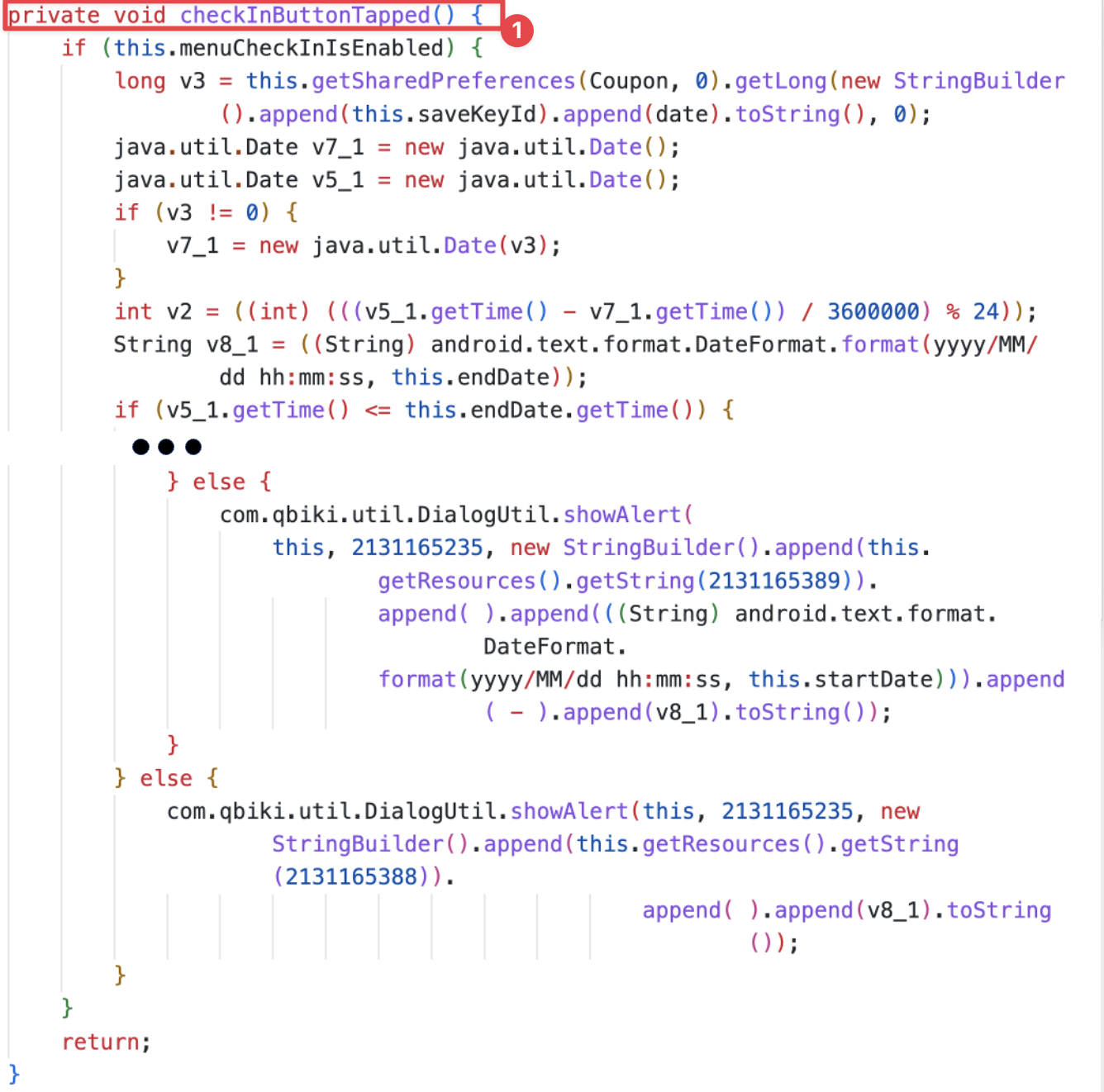}}
    \caption{Too few highlights for large code samples do not help.}
    \label{fig:result-rq-3.1-long-bad}
\end{figure}

\vfill\eject

\section{RQ 3.2: Accuracy of Localization -- Examples}
\label{subsec:appendix-rq-3.2}

\begin{figure}[htbp]
    \centering
    \fbox{\includegraphics[width=0.475\textwidth]{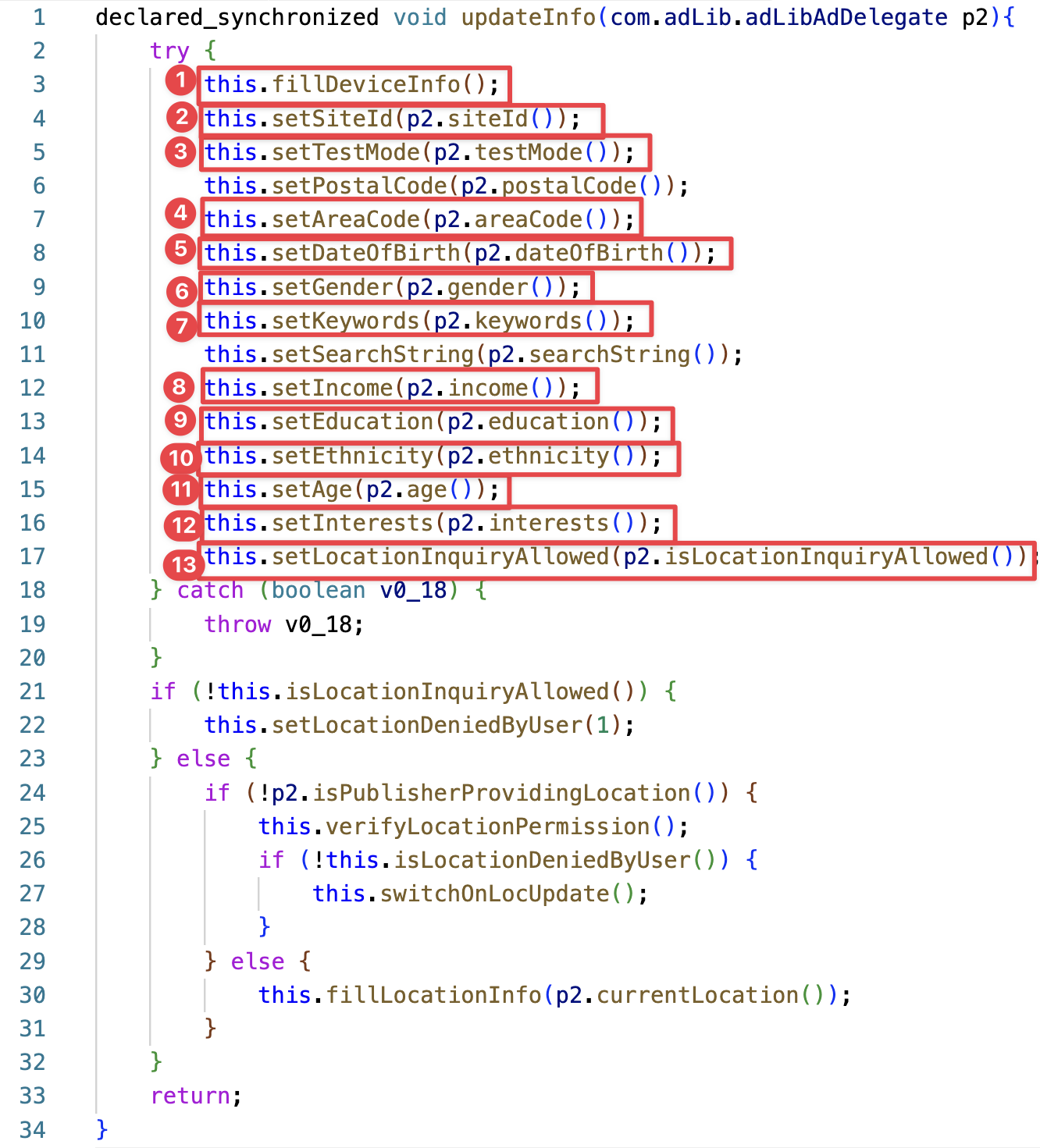}}
    \caption{Agreement analysis: code snippet 2}
    \label{fig:result-rq3-2}
\end{figure}

\newpage

\end{document}